\documentclass[sigconf]{acmart}

\AtBeginDocument{%
  }

\copyrightyear{2025}
\acmYear{2025}
\setcopyright{cc}
\setcctype{by}
\acmConference[RecSys '25]{Proceedings of the Nineteenth ACM Conference on Recommender Systems}{September 22--26, 2025}{Prague, Czech Republic}
\acmBooktitle{Proceedings of the Nineteenth ACM Conference on Recommender Systems (RecSys '25), September 22--26, 2025, Prague, Czech Republic}\acmDOI{10.1145/3705328.3748068}
\acmISBN{979-8-4007-1364-4/2025/09}

\usepackage{graphicx}
\usepackage{subcaption}
\usepackage{multicol}
\usepackage{algorithm}
\usepackage{algorithmic}
\usepackage{enumitem}
\usepackage{xcolor} 
\definecolor{lightblue}{rgb}{0.2, 0.6, 1} 
\usepackage{multirow} 

\begin{document}

\title{$R^{4}$ec: A Reasoning, Reflection, and Refinement Framework for Recommendation Systems}

\author{Hao Gu}
\authornote{Both authors contributed equally to this research.}
\affiliation{%
  \institution{Institute of Automation, \\ Chinese Academy of Sciences}
  \city{Beijing}
  \country{China}
}
\email{guhao22@ia.ac.cn}

\author{Rui Zhong}
\authornotemark[1]
\affiliation{%
  \institution{Kuaishou Technology}
  \city{Beijing}
  \country{China}}
\email{zhongrui@kuaishou.com}

\author{Yu Xia}
\affiliation{%
  \institution{University of Chinese Academy \\ of Sciences}
  \city{Beijing}
  \country{China}}
\email{xiayu24@mails.ucas.ac.cn}

\author{Wei	Yang}
\affiliation{%
  \institution{Kuaishou Technology}
  \city{Beijing}
  \country{China}}
\email{yangwei08@kuaishou.com}

\author{Chi Lu}
\affiliation{%
  \institution{Kuaishou Technology}
  \city{Beijing}
  \country{China}}
\email{luchi@kuaishou.com}

\author{Peng Jiang}
\affiliation{%
  \institution{Kuaishou Technology}
  \city{Beijing}
  \country{China}}
\email{jiangpeng@kuaishou.com}

\author{Kun Gai}
\affiliation{%
  \institution{Kuaishou Technology}
  \city{Beijing}
  \country{China}}
\email{yuyue06@kuaishou.com}

\renewcommand{\shortauthors}{Hao Gu et al.}

\begin{abstract}
  Harnessing Large Language Models (LLMs) for recommendation systems has emerged as a prominent avenue, drawing substantial research interest. However, existing approaches primarily involve basic prompt techniques for knowledge acquisition, which resemble System-1 thinking.  This makes these methods highly sensitive to errors in the reasoning path, where even a small mistake can lead to an incorrect inference.  To this end, in this paper, we propose $R^{4}$ec, a reasoning, reflection and refinement framework that evolves the recommendation system into a weak System-2 model.  Specifically, we introduce two models: an actor model that engages in reasoning, and a reflection model that judges these responses and provides valuable feedback. Then the actor model will refine its response based on the feedback, ultimately leading to improved responses.  We employ an iterative reflection and refinement process, enabling LLMs to facilitate slow and deliberate System-2-like thinking.  Ultimately, the final refined knowledge will be incorporated into a recommendation backbone for prediction.  We conduct extensive experiments on Amazon-Book and MovieLens-1M datasets to demonstrate the superiority of $R^{4}$ec.  We also deploy $R^{4}$ec on a large scale online advertising platform, showing 2.2\% increase of revenue. Furthermore, we investigate the scaling properties of the actor model and reflection model. 
  
\end{abstract}




\maketitle

\section{Introduction}
Nowadays, recommendation systems play a vital role in various online applications to alleviate the information overload problem and fulfill the information needs of users \cite{zhao2023recommender,wang2024rethinking,bai2024finetuning,gu2025mathcal,xia2025hierarchical}. Besides, Large Language Models (LLMs) have achieved remarkable breakthroughs in Natural Language Processing (NLP), demonstrating impressive capacity in natural language understanding and text generation \cite{touvron2023llama,achiam2023gpt,ouyang2022training,bai2023qwen}.  Consequently, LLM-enhanced recommendation systems have received much attention and have been actively explored currently \cite{liao2023llara,bao2023tallrec,xi2024towards,xu2024prompting}. 


At the core of integrating LLMs with recommendation systems is to harness LLM's extensive open-world knowledge and impressive reasoning capabilities to benefit recommendation systems.  Initial attempts \cite{geng2022recommendation,wang2023zero,gao2023chat} has employed in-context learning to align LLMs with recommendation problems. They employ LLMs to rerank the candidate items filtered by traditional models (such as MF \cite{koren2009matrix} and LightGCN \cite{he2020lightgcn}).  However, these approaches fail to achieve satisfactory performance. Most recently, KAR \cite{xi2024towards} leverages chain-of-thought prompt technique, which facilitates the LLMs to break down recommendation tasks into a series of intermediate steps and generate the knowledge of LLMs regarding user preference and factual knowledge on items in a step-by-step manner. Then the extracted knowledge is treated as additional input features for downstream recommendation backbone.  Although existing LLM-based methods have achieved remarkable success, numerous challenges hinders their real-world applications. 

On the one hand, they are sensitive to mistakes in the reasoning path \cite{wu2024large,ji2025test,xi2024enhancing}, which means any mistake can lead to an incorrect answer. These shortcomings are attributed to the model’s reliance on fast, intuitive System-1 thinking \cite{kahneman2011thinking}, which responds directly based on internally encoded perceptual information and world knowledge \cite{kahneman2011thinking,ji2025test}.  To address this issue, inspired by the procedure of human cognition \cite{hegel1991encyclopaedia}, We introduce an iterative reflection and refinement mechanism, facilitating slow and deliberate System-2 thinking. Specifically, we incorporate two models: an actor model and a reflection model. The actor model is capable of iteratively refining its responses based on feedback from the reflection model, shifting LLMs from System-1 thinking to a weak System-2 thinking.





On the other hand, these methods \cite{xi2024towards,xu2024prompting} often necessitate numerous API calls, such as GPT-3.5 \cite{achiam2023gpt}, leading to exorbitant inference latency and financial costs. This is typically intolerable in practical applications.  Accordingly,  we develop small LLMs like Qwen-2.5 7B \cite{yang2024qwen2} for user preference and item factual knowledge acquisition instead.

In this paper, we dive into how to employ System-2 thinking with small LLMs for recommendation tasks.  To this end, we propose $\mathcal{R}^{4}$ec, a reasoning, reflection and refinement framework for enhancing recommendation system with LLMs. Specifically, we aim to train smaller LLMs to develop capabilities in reasoning, reflection, and refinement for user and item, respectively.  As indicated in \cite{xi2024enhancing}, the effectiveness of these mechanisms can be limited by factors such as model's capacity to accurately assess its own response. Thus, we employ a two-role paradigm, introducing two models: the actor model $\pi_{\theta}$ and the reflection model $\pi_{\psi}$.  The actor model $\pi_{\theta}$ is tasked with reasoning about user preference or item factual knowledge and refining this knowledge based on feedback.  In contrast, the reflection model $\pi_{\psi}$ judges the rationality of the actor model's output and provides reflections for those outputs deemed unreasonable. The design philosophy of our approach is to enable the reflection and actor models to iteratively reflect and refine the knowledge until no further errors are detected, i.e., until the reflection model deems the knowledge to be rational.  This process takes a step towards System-2 thinking, where deliberate and reflective reasoning is applied. Notably, we deploy distinct actor and reflection models for user preference and item factual knowledge respectively.  Ultimately, the final refined user preference and item factual knowledge will be incorporated into the a recommendation backbone for prediction.

In summary, our main contributions are:
\begin{itemize}[leftmargin=*]
    \item We introduce $\mathcal{R}^{4}$ec framework, the first study within recommendation systems to explore System-2 thinking through iterative reflection and refinement mechanism, demonstrating its significant potential in the recommendation domain.
    \item We advance the reflection and refinement mechanisms in LLMs, transforming them from intuitive System-1 thinking to deliberate System-2 reasoning.  We achieve this by incorporating an actor model that learns reasoning and refinement capabilities, and a reflection model that develops reflection abilities.
    \item We conduct comprehensive experiments on two public datasets and online industrial experiments to demonstrate the effectiveness of $\mathcal{R}^{4}$ec.  Additionally, we shed light on the scaling properties of the actor and the reflection model.
\end{itemize}

\section{Related Work}
\subsection{LLMs for Recommendation}
Recently, leveraging Large Language Models (LLMs) for recommendation systems has attracted considerable attention \cite{lin2024bridging,liu2023chatgpt}.  Generally, current LLM empowered recommenders can be primarily categorized into two main trends based on the distinct roles that LLMs play within the recommendation pipeline \cite{zhao2023recommender,xu2024prompting}.  (1) \textit{LLM as a ranker}.  This paradigm typically employs a frozen LLM to generate a ranked list that aligns with user interests \cite{wang2024learnable,dai2023uncovering,wang2023zero}.  However, the recommendation capabilities are somewhat limited due to the inherent gap between the LLMs' pre-training and recommendation tasks.  Therefore,  recent studies \cite{bao2023tallrec,zhang2023recommendation} utilize instruction tuning for LLMs to inject recommendation capabilities. TallRec \cite{bao2023tallrec} fine-tunes LLAMA-7B \cite{touvron2023llama} using LoRA \cite{hu2021lora}, a parameter-efficient approach, in a two-stage tuning process. The first stage utilizes general data from Alpaca \cite{taori2023stanford}, followed by the second stage with recommendation data.  (2) \textit{LLM as a knowledge enhancer}. This method mainly utilizes LLMs to generate auxiliary information to improve the performance of recommendation models \cite{xu2024prompting}.  For example, \cite{du2024enhancing} proposed a job recommendation model that leverages the summarization capabilities of LLMs to extract user job requirements.  Similarly, KAR \cite{xi2024towards} leverages factual knowledge stored in LLMs to enhance movie and book recommendation.


\subsection{Self-refine in LLMs}
Self-refine, which aims to enhance the quality of responses from large language models (LLMs) by refining them using LLMs, have demonstrated effectiveness in various reasoning tasks \cite{wu2024large,madaan2024self}. These tasks include arithmetic reasoning \cite{welleck2022generating,madaan2024self,kumar2024training}, open-domain question answering \cite{dai2023uncovering,yu2023improving}, code generation \cite{chen2023teaching,jiang2024training} and others.

The self-refine methods can be categorized into three groups based on the source of feedback. (1) \textit{Intrinsic}. Intrinsic self-refine methods instruct LLMs to generate feedback on their own responses and correct them without external feedback.  However, recent studies \cite{olausson2023self} report that intrinsic self-refine does not improve or even degrade performances. (2) \textit{External Tools.} External tools employed for self-refine encompass code interpreters, which facilitate code generation tasks \cite{chen2023teaching}, and symbolic reasoners, applied to arithmetic reasoning \cite{pan2023logic}. Additionally, \cite{gou2023critic} builds a web search tool based on Google to retrieve information for validating correctness of the response.  (3) \textit{Fine-tuning.} This line of work improves reflection ability of open-source LLM by fine-tuning them on reflection datasets generated by GPT-4 \cite{ke2023critiquellm,kim2024language,li2023generative}. For example, \cite{paul2023refiner} first defines multiple error types for natural language reasoning tasks and develops specific feedback templates accordingly.  Then they train a reflection generation model using synthetic feedback data.  Our work falls under the third category, which involves fine-tuning reflection model to generate feedback given initial responses and fine-tuning refinement model to generate refined answers given the initial responses and feedback.

\section{Method}

\subsection{Overview} \label{dataset construction}

Our proposed $\mathcal{R}^{4}$ec utilizes three primary capabilities, reasoning, reflection, and refinement, to develop System-2 thinking for recommendation. We formally define the key concepts:
\begin{itemize}[leftmargin=*]
    \item Reasoning capability refers to generating a response for a given question. In recommendation tasks, this involves inferring user preferences based on their interaction history for example.
    \item Reflection capability denotes the ability of LLMs to identify flaws in initial responses and to offer feedback for correction.
    \item Refinement capability involves generating a refined response based on initial responses and feedback.
\end{itemize}

We will introduce two small LLMs: an actor model $\pi_{\theta}$ for reasoning and refinement, and a reflection model $\pi_{\psi}$ for reflection capability.  Our primary motivation is to enable the actor model $\pi_{\theta}$ continuously refine its generated knowledge under the guidance of the reflection model $\pi_{\psi}$. This process facilitates the extraction of user preferences and item factual knowledge, gradually evolving into a more deliberate System-2 thinking approach. The final refined knowledge is subsequently integrated into a traditional recommendation backbone for prediction.

\begin{figure}[ht]  
\centering  
\includegraphics[width=0.3\textwidth]{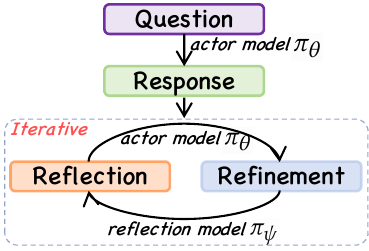
} 
\caption{Our iterative reflection and refinement mechanism.}  
\label{fig:scaling properties of actor models}  
\end{figure}



\begin{figure*}[ht]  
\centering  
\includegraphics[width=0.85\textwidth]{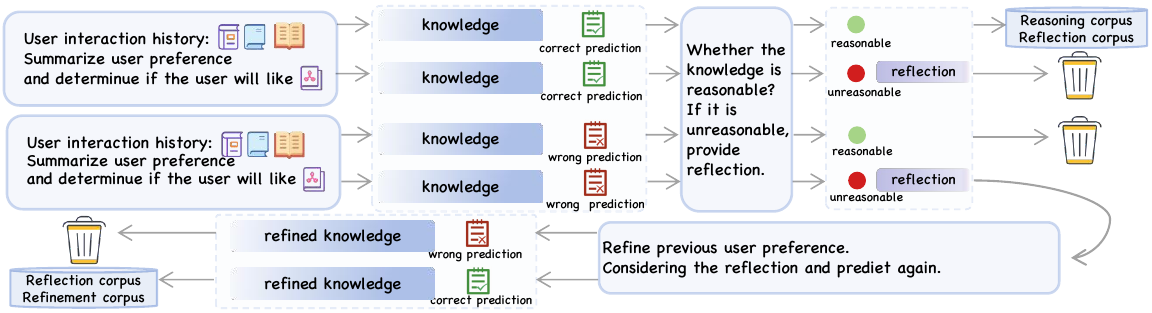} 
\caption{Overview of our user preference reasoning, reflection and refinement dataset construction process.}  
\label{fig:overview}  
\end{figure*}

\subsection{User Preference Dataset for $\mathcal{R}^4$ec}\label{user preference for r4ec}
In a user preference inference task, given the interaction history and corresponding rating of a user, the model is required to summarize user preference.  To construct reasoning, reflection and refinement dataset for user preference, we start from the widely available recommendation data $\{ hist_{k}, item_{k},label_{k} \}_{k=1}^{\left|D\right|}$, where $hist_{k}$, $item_{k}$ represent the interaction history of user $u_{k}$ and the target item respectively. $label_{k}$ represents whether user $u_{k}$ will like $item_{k}$, which corresponds to the gold answer.

We adopt a \textit{reasoning while predicting} paradigm, which leverages the reasoning capabilities of LLMs to to provide predictions and justifications. Specifically, for a user $u$, given his interaction history $hist$ and the target item $item$, we prompt LLM $\mathcal M$ using the \textbf{\textit{User Preference Reasoning Construction Prompt}} $\mathcal{P}_{reason}^{user}$ to predict whether the user $u$ will like the target item $item$, along with rationales to justify the prediction. These rationales are natural language explanations that provide supporting reasoning for the prediction, which we set in the prompt as the user preference knowledge.  Formally, this process can be formulated as:
\begin{equation}
    <u_{pre},pred> \leftarrow \mathcal M(\mathcal{P}_{reason}^{user}(hist,item)) 
\end{equation}
where $u_{pre}$ and $pred$ denotes user preference knowledge and the prediction indicating whether the user will like the item.

Upon failing to solve a problem, human students typically analyze the errors in their solutions and reflect on how to correct them. Inspired by this process, we design the reflection mechanism that enables the model to acquire the capability to identify flaws in the responses. Specifically, we prompt LLM $\mathcal M$ using the \textbf{\textit{User Preference Reflection Construction Prompt}} $\mathcal{P}_{reflect}^{user}$. We task the LLM with judging whether the user preference knowledge $u_{pre}$ is reasonable. For unreasonable user preference knowledge, $\mathcal M$ will generate reflections that not only pinpoint flaws in the responses but also provide valuable suggestions for correction.  Formally, this process can be formulated as:
\begin{equation}
    <judge^{u},reflect^{u}> \leftarrow \mathcal M(\mathcal{P}_{reflection}^{user}(hist,item,u_{pre})) 
\end{equation}
Here, $judge^{u}$ represents the assessment of whether the inferred user preference knowledge $u_{pre}$ is reasonable, while $reflect^{u}$ refers to the reflection on $u_{pre}$. Notably, if LLM $\mathcal M$ deems $u_{pre}$ to be reasonable, $reflect^{u}$ will be empty.  

Given that we have access to the user's true preference for items, which we denote as $label$. Samples that lead to correct predictions and are deemed reasonable by $\mathcal M$ will be incorporated into the User Preference Reasoning Dataset $\mathcal{D}_{reason}^{u}$. Additionally, these samples will be included as positive instances in the User Preference Reflection Dataset $\mathcal{D}_{reflect}^{u}$.  For user preference that lead to correct prediction but are deemed unreasonable by the $\mathcal{M}$, as well as those that lead to incorrect predictions but are considered reasonable, there is a high likelihood of issues in the reasoning or reflection process. Consequently, we discard these samples.

Finally, for samples that lead to incorrect predictions and are concurrently considered unreasonable by $\mathcal{M}$, we leverage these instances to develop a dataset specifically aimed at enhancing the model's refinement capabilities. To this end, we introduce the \textbf{User Preference Refine Construction Prompt} $\mathcal{P}_{refine}^{user}$, then engage $\mathcal{M}$ to generate a new prediction along with the refined user preference. This process can be represented as:

\begin{equation}
    <u_{pre}^{r}, pred'> \leftarrow \mathcal{M}(\mathcal{P}_{refine}^{user}(hist,item,u_{pre},reflect^u))
\end{equation}
where $u_{pre}^{r}$ denotes the refined user preference knowledge and $pred'$ represents the new prediction.

For samples where the new prediction $pred'$ matches the label, this indicates that not only the reflection successfully identifies errors in original user preference, but also the subsequent refinement yields the correct answer. We consider such reflections effective and will incorporate $s_{reflect}^{u}=\{(hist,u_{pre}),(judge^{u},reflect^{u})\}$  into the User Preference
Reflection Dataset $\mathcal{D}_{reflect}^{u}$. Additionally, sample $\{(hist,u_{pre},reflect^{u}),(u_{pre}^{r})\}$ will be added to the User Preference Refinement Dataset: $\mathcal{D}_{refine}^{u}$.


\subsection{Item Factual Dataset for $\mathcal{R}^4$ec}\label{item knowledge for r4ec}
To construct the dataset capable of endowing models with the abilities to reason, reflect, and refine upon item factual knowledge, we require supervision signals. However, these signals are not as intuitive as those found in the user preference data. To this end, we start from the following dataset structure $\{item_{k},pos_{k},neg_{k},tar_{k},label_{k}\}_{k=1}^{\left|D'\right|}$, where $item_{k}$ denotes the information of the target item $i_{k}$. $pos_{k}$ and $neg_{k}$ represent the interaction history of the users who like and dislike $i_{k}$, while $tar_{k}$ denotes the interaction history of the target user.  $label_{k}$ indicates whether the target user will like $i_{k}$, serving as the gold answer, i.e. supervision signal.

Building on the methodology used for constructing user preference datasets with capabilities for reasoning, reflection, and refinement, we apply a similar approach to develop the corresponding datasets for items.  Suppose $s_{reason}^{i}$, $s_{reflect}^{i}$, $s_{refine}^{i}$ denote sample from the item factual reasoning dataset $\mathcal{D}_{reason}^{i}$, item factual reflection dataset $\mathcal{D}_{reflect}^{i}$ and item factual refinement dataset $\mathcal{D}_{refine}^{i}$. Then, we can represent the obtained dataset as follows:
\begin{align}
s_{reason}^{i} &= \{(item,pos,neg),(i_{fact})\} \\
s_{reflect}^{i} &= \{(item,pos,neg,i_{fact}),(judge^{i},reflect^{i})\} \\
s_{refine}^{i} &= \{(item,pos,neg,i_{fact},reflect^{i}),(i_{fact}^{r})\}
\end{align}

In this context, $i_{fact}$, $judge^{i}$, and $reflect^{i}$ correspond to the item factual knowledge, the assessment of whether this item factual knowledge is reasonable, and the reflection on this knowledge. Finally, $i_{fact}^{r}$ is the refined item factual knowledge.

\begin{algorithm}
    \caption{User Preference Dataset Construction for $\mathcal{R}^{4}$ec}
    \label{algorithm}
    \renewcommand{\algorithmicrequire}{\textbf{Input:}}
\renewcommand{\algorithmicensure}{\textbf{Output:}}
    \begin{algorithmic}[1]
        \REQUIRE{Sample set $\mathcal{D} = \{(hist_{k},item_{k},label_{k})_{k=1}^{\left|D\right|}\}$, LLM $\mathcal{M}$, reasoning construction prompt template $\mathcal{P}_{reason}^{user}$, reflection construction prompt template $\mathcal{P}_{reflect}^{user}$ and refinement construction prompt template $\mathcal{P}_{refine}^{user}$ for user preference.}
        \ENSURE{user preference reasoning dataset: $\mathcal{D}_{reason}^{u}$, user preference reflection dataset: $\mathcal{D}_{reflect}^{u}$ and user preference refinement dataset: $\mathcal{D}_{refine}^{u}$}
        \STATE{Initialize $\mathcal{D}_{reason}^{u} = \mathcal{D}_{reflect}^{u} = \mathcal{D}_{refine}^{u} = \emptyset$}
        \FOR{each $(hist,item,label) \in \mathcal{D}$}
            \STATE{$<u_{pre},pred> \leftarrow \mathcal{M}(\mathcal{P}_{reason}^{user}(hist,item))$}
            \STATE{$<judge^{u},reflect^{u}> \leftarrow \mathcal{M} (\mathcal{P}_{reflect}^{user}(hist,item,u_{pre})$}
\IF{$\text{judge}^{u} == \text{"the user preference is reasonable"} \ \AND$  \\
        \hspace{1em} $\text{pred} = \text{label}$}
                \STATE $\mathcal{D}_{reason}^{u} += \{(hist),(u_{pre})\}$
                \STATE $\mathcal{D}_{reflect}^{u}  += \{(hist,u_{pre}),(judge^{u},reflect^{u}=\text{""})\}$
\ELSIF{$\text{judge}^{u} == \text{"the user preference is not reasonable"}$ \\
        \hspace{1em} $ \AND \  \text{pred} \neq \text{label}$}
                \STATE $<u_{pre}^{r},pred'>\leftarrow \mathcal{M}(\mathcal{P}_{refine}^{user}(hist,item,u_{pre},reflect^u))$
                \IF{$pred' = label$}
            \STATE $\mathcal{D}_{reflect}^{u} += \{(hist,u_{pre}),(judge^{u},reflect^{u})\}$
            \STATE $\mathcal{D}_{refine}^{u} += \{(hist,u_{pre},reflect^{u}),(u_{pre}^{r})\}$  
            \ENDIF
            \ELSE
                \STATE $Discard (hist,item,label)$
            \ENDIF
        \ENDFOR
    \end{algorithmic}
\end{algorithm}

\subsection{Training actor and reflection model}
In Sec. \ref{user preference for r4ec} and Sec. \ref{item knowledge for r4ec}, we describe how to construct the user preference and item factual datasets for $\mathcal{R}^{4}$ec, resulting in the datasets that enhance LLM's reasoning, reflection, and refinement capabilities for both users and items.  In this section, we will explain how to train the actor and reflection models for users and items.  

Without loss of generality, we denote $\mathcal{D}_{reason}$, $\mathcal{D}_{reflect}$, $\mathcal{D}_{refine}$ as the datasets that endow the models with reasoning, reflection, and refinement capabilities. Actor model and reflection model are denoted as $\pi_{\theta}$ and $\pi_{\psi}$.

First, we will train the actor model $\pi_{\theta}$ with basic reasoning and refinement ability.  Refinement capability enables the generation of a revised response based on the problem, the previous response, and reflections on that response.  In scenarios where both the previous response and its reflection are absent, this capability degenerates to basic reasoning.  Recognizing this, we employ $\mathcal{D}_{reason}$ and $\mathcal{D}_{refine}$ to equip actor model $\pi_{\theta}$ with both reasoning and refinement capabilities.  The loss of training $\pi_{\theta}$, i.e. $\mathcal{L}_{actor}$, is as follows:
\begin{equation}
\mathcal{L}_{reason} = \mathbb{E}_{(x,y) \sim \mathcal{D}_{reason}} \left[ \log \pi_{\theta}(y|x) \right] 
\end{equation}

\begin{equation}
\mathcal{L}_{refine} = \mathbb{E}_{(x',y') \sim \mathcal{D}_{refine}} \left[ \log \pi_{\theta}(y'|x') \right]
\end{equation}

\begin{equation}
\mathcal{L}_{actor} = \mathcal{L}_{reason} + \mathcal{L}_{refine}
\end{equation}

Here, $(x,y)$ denotes sample from $\mathcal{D}_{reason}$ and $(x',y')$ denotes sample from $\mathcal{D}_{refine}$. Loss functions $\mathcal{L}_{reason}$ and $\mathcal{L}_{refine}$ correspond to the reasoning and refinement capabilities, respectively.

Next, we equip the reflection model $\pi_{\psi}$ with the reflection capability.  Specifically, we train $\pi_{\psi}$ through supervised fine-tuning with the collected reflection dataset $\mathcal{D}_{reflect}$. Suppose $(x^*,y^*)$ represents sample from $\mathcal{D}_{reflect}$.  Then the loss $\mathcal{L}_{reflect}$ for the reflection model $\pi_{\psi}$ is as follows:

\begin{equation}
\mathcal{L}_{reflect} = \mathbb{E}_{(x^*,y^*) \sim \mathcal{D}_{reflect}} \left[ \log \pi_{\psi}(y^*|x^*) \right]
\end{equation}

In this way, we can obtain a reflection model that can provide constructive feedback on the reasoning paths of the actor model.  Additionally, fine-tuning all parameters of the LLMs is time-consuming and resource-intensive \cite{liao2023llara,bao2023tallrec}. Thus, we employ the LoRA technique \cite{hu2021lora} to reduce computational demands while maintaining good performance.



\begin{figure*}[ht]  
\centering  
\includegraphics[width=0.84\textwidth]{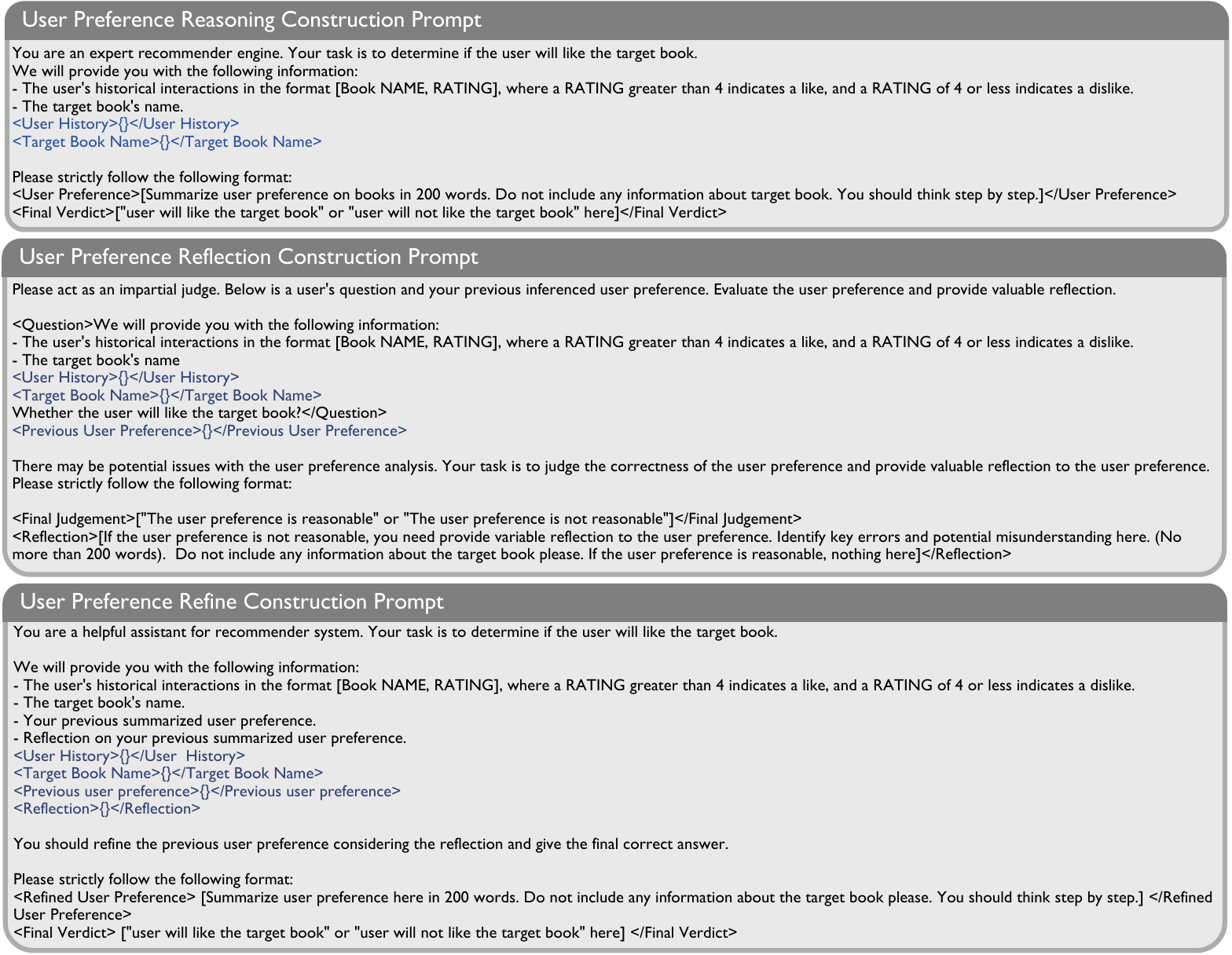} 
\caption{Prompt template for constructing user preference reasoning, reflection and refinement dataset on Amazon-Book.}  
\label{fig:prompt}  
\end{figure*}


\subsection{Inference Strategy} \label{Inference Strategy}

To effectively utilize our learned reasoning, reflection, and refinement capabilities, we implement two distinct inference strategies: (1): Iterative Refinement and (2): Reflection as a Filter Function.\cite{xi2024enhancing,zheng2024critic,snell2024scaling,qu2024recursive}. Next, we will demonstrate these strategies through user preference inference problem, while the strategy maintains equivalent applicability to item factual knowledge inference.

\paragraph{Iterative Refinement} This strategy is formalized as an iterative optimization process \cite{shinn2024reflexion}. In the first trial, actor model $\pi_{\theta}$ will generate user preference knowledge. If the reflection model $\pi_{\psi}$ identifies flaws in response, $\pi_{\theta}$ will incorporate the reflection from $\pi_{\psi}$ to produce refined user preference knowledge. However, the refined user preference knowledge may still contain errors. Therefore, we can iteratively inspect the refined knowledge and refine it further if errors are identified. The final knowledge is only produced when it satisfies $\pi_{\psi}$ (i.e., when $\pi_{\psi}$ considers the user preference knowledge to be reasonable) or when the maximum number of retries is reached.  The iterative reflection and refinement mechanism evolves our knowledge acquisition into a slow and deliberate System-2 thinking.

\paragraph{Reflection as a Filter} As indicated in previous studies \cite{wang2022self,zheng2024critic,xi2024enhancing}, self-consistency technique is an effective approach for improving accuracy. Thus, we can generate several user preference knowledge via actor model $\pi_{\theta}$, then the reflection model $\pi_{\psi}$ is employed to filter out unreasonable user preference. The final user preference knowledge embedding is obtained by averaging the filtered preference. If no user preferences are deemed reasonable by $\pi_{\psi}$, we average all the available knowledge instead. 
\subsection{Knowledge Utilization}
After we derive the final user preference knowledge $u_{pre}$ and item factual knowledge $i_{fact}$ from the actor and reflection model through the iterative refinement inference strategy. We need to transform the generated text-based knowledge into dense vectors.  Specifically, we employ a knowledge encoder $\mathcal{E}ncoder$, such as
BGE-M3 \cite{chen2024bge}:


\begin{equation}
e^u = \mathcal{E}ncoder(u_{pre}),  e^i = \mathcal{E}ncoder(i_{fact})
\end{equation}

Here $e^{u}$ and $e^{i}$ denote the dense representations of user preference knowledge and item factual knowledge, respectively.

The recommendation task is generally formulated as a binary classification problem.  Generally, we estimate the click probability as follows:

\begin{equation}
    \hat{y} = \mathcal{M}(x,\mathcal{F}_{u}(e^u),\mathcal{F}_{i}(e^i))
\end{equation}

Here x represents categorical features for conventional recommendation systems.  $\mathcal{M}$ represents the recommendation backbones, and $\mathcal{F}_{u}$ and $\mathcal{F}_{i}$ denote the connector for user preference knowledge and item factual knowledge, implemented as a MLP, respectively. During training, $\mathcal{F}_{i}$, $\mathcal{F}_{u}$ and $\mathcal{M}$ are jointly optimized via the binary cross-entropy loss. 


\section{Experiments}

\subsection{Experimental Setup} \label{experimental setup}
\subsubsection{Datasets.}

\begin{table}
\renewcommand{\arraystretch}{0.8}
  \caption{Statistics of datasets used in this paper.}
  \label{tab:statistics}
  \begin{tabular}{l|c|c|c}
    \toprule
    Dataset & Users & Items & Interactions \\
    \midrule
    Amazon-Book & 11906 & 17332 & 1.4 million \\
    MovieLens-1M & 6040 & 3706 & 1 million \\
    Industrial Dataset & 0.4 billion & 10 million & 2.3 billion \\
    \bottomrule
  \end{tabular}
\end{table}

\begin{table*}[t] 
\centering
\renewcommand{\arraystretch}{0.85}
\begin{tabular}{ccc|cccccccc}
\toprule
\multirow{2}{*}{Backbones} & \multirow{2}{*}{Method} & \multirow{2}{*}{LLM} & \multicolumn{4}{c}{Amazon-Book} & \multicolumn{4}{c}{MovieLens-1M } \\
\cmidrule(lr){4-7} \cmidrule(lr){8-11} 
 & & & \textbf{AUC} & \textbf{Rel. Impr.} & \textbf{LogLoss} & \textbf{Rel. Impr.} & \textbf{AUC} & \textbf{Rel. Impr.} & \textbf{LogLoss} & \textbf{Rel. Impr.} \\
\midrule
\multirow{4}{*}{DIEN \cite{zhou2019deep}} & Base & \rule[0.5ex]{0.6cm}{0.4pt}
 & 0.8280 & \rule[0.5ex]{0.6cm}{0.4pt} & 0.5004 & \rule[0.5ex]{0.6cm}{0.4pt} & 0.7755 & \rule[0.5ex]{0.6cm}{0.4pt} & 0.5600 & \rule[0.5ex]{0.6cm}{0.4pt}\\
 & KAR & GPT-3.5
 & 0.8360 & \textcolor{red!60}{$\uparrow$\,0.97\%} & 0.4872 & \textcolor{blue!60}{$\downarrow$\,2.64\%} & 0.7938 & \textcolor{red!60}{$\uparrow$\,2.35\%}& 0.5406 & \textcolor{blue!60}{$\downarrow$\,3.46\%}\\
 & $\mathcal{R}^2$ec & Qwen2.5-7B
 & 0.8434 & \textcolor{red!60}{$\uparrow$\,1.86\%} & 0.4827 & \textcolor{blue!60}{$\downarrow$\,3.53\%} & 0.7963 & \textcolor{red!60}{$\uparrow$\,2.68\%}& 0.5382 & \textcolor{blue!60}{$\downarrow$\,3.89\%} \\
 & $\mathcal{R}^4$ec & Qwen2.5-7B
 & 0.8488 & \textbf{\textcolor{red}{$\uparrow$\,2.51\%}} & 0.4699 & \textbf{\textcolor{blue}{$\downarrow$\,6.09\%}}& 0.8006 & \textbf{\textcolor{red}{$\uparrow$\,3.23\%}}& 0.5348 & \textbf{\textcolor{blue}{$\downarrow$\,4.50\%}}\\
 \midrule
 \multirow{4}{*}{GRU4Rec \cite{hidasi2015session}} & Base & \rule[0.5ex]{0.6cm}{0.4pt}
 & 0.8281 & \rule[0.5ex]{0.6cm}{0.4pt} & 0.4992 & \rule[0.5ex]{0.6cm}{0.4pt} & 0.7760 & \rule[0.5ex]{0.6cm}{0.4pt} & 0.5589 & \rule[0.5ex]{0.6cm}{0.4pt}\\
 & KAR & GPT-3.5
 & 0.8376 & \textcolor{red!60}{$\uparrow$\,1.15\%} & 0.4915 & \textcolor{blue!60}{$\downarrow$\,1.54\%} & 0.7942 & \textcolor{red!60}{$\uparrow$\,2.34\%}& 0.5401 & \textcolor{blue!60}{$\downarrow$\,3.36\%}\\
 & $\mathcal{R}^2$ec & Qwen2.5-7B
 & 0.8410 & \textcolor{red!60}{$\uparrow$\,1.56\%} & 0.4825 & \textcolor{blue!60}{$\downarrow$\,3.35\%} & 0.7955 & \textcolor{red!60}{$\uparrow$\,2.51\%} & 0.5407 & \textcolor{blue!60}{$\downarrow$\,3.25\%}\\
 & $\mathcal{R}^4$ec & Qwen2.5-7B
 & 0.8492 & \textbf{\textcolor{red}{$\uparrow$\,2.55\%}} & 0.4690 & \textbf{\textcolor{blue}{$\downarrow$\,6.05\%}} & 0.8002 & \textbf{\textcolor{red}{$\uparrow$\,3.12\%}} & 0.5370 & \textbf{\textcolor{blue}{$\downarrow$\,3.92\%}}\\
 \midrule
 \multirow{4}{*}{AutoInt \cite{song2019autoint}} & Base & \rule[0.5ex]{0.6cm}{0.4pt}
 & 0.8261 & \rule[0.5ex]{0.6cm}{0.4pt} & 0.5007 & \rule[0.5ex]{0.6cm}{0.4pt} & 0.7736 & \rule[0.5ex]{0.6cm}{0.4pt}& 0.5618 & \rule[0.5ex]{0.6cm}{0.4pt}\\
 & KAR & GPT-3.5
 & 0.8404 & \textcolor{red!60}{$\uparrow$\,1.73\%} & 0.4842 & \textcolor{blue!60}{$\downarrow$\,3.29\%} & 0.7949 & \textcolor{red!60}{$\uparrow$\,2.75\%} & 0.5419 & \textcolor{blue!60}{$\downarrow$\,3.54\%} \\
 & $\mathcal{R}^2$ec & Qwen2.5-7B
 & 0.8448 & \textcolor{red!60}{$\uparrow$\,2.26\%} & 0.4755 & \textcolor{blue!60}{$\downarrow$\,5.03\%} & 0.7952 & \textcolor{red!60}{$\uparrow$\,2.79\%} & 0.5386 & \textcolor{blue!60}{$\downarrow$\,4.12\%} \\
 & $\mathcal{R}^4$ec & Qwen2.5-7B
 & 0.8494 & \textbf{\textcolor{red}{$\uparrow$\,2.82\%}} & 0.4686 & \textbf{\textcolor{blue}{$\downarrow$\,6.41\%}}& 0.8008 & \textbf{\textcolor{red}{$\uparrow$\,3.52\%}} & 0.5347 & \textbf{\textcolor{blue}{$\downarrow$\,4.82\%}} \\
\midrule
\multirow{4}{*}{FiGNN \cite{li2019fi}} & Base & \rule[0.5ex]{0.6cm}{0.4pt}
 & 0.8273 & \rule[0.5ex]{0.6cm}{0.4pt} & 0.4993 & \rule[0.5ex]{0.6cm}{0.4pt}& 0.7742 & \rule[0.5ex]{0.6cm}{0.4pt}& 0.5611 & \rule[0.5ex]{0.6cm}{0.4pt}\\
 & KAR & GPT-3.5
 & 0.8393 & \textcolor{red!60}{$\uparrow$\,1.45\%} & 0.4826 & \textcolor{blue!60}{$\downarrow$\,3.34\%} & 0.7947 & \textcolor{red!60}{$\uparrow$\,2.65\%} & 0.5422 & \textcolor{blue!60}{$\downarrow$\,3.37\%}\\
 & $\mathcal{R}^2$ec & Qwen2.5-7B
 & 0.8452 & \textcolor{red!60}{$\uparrow$\,2.16\%} & 0.4752 & \textcolor{blue!60}{$\downarrow$\,4.83\%} & 0.7968 & \textcolor{red!60}{$\uparrow$\,2.92\%} & 0.5374 & \textcolor{blue!60}{$\downarrow$\,4.22\%}\\
 & $\mathcal{R}^4$ec & Qwen2.5-7B
 & 0.8495 & \textbf{\textcolor{red}{$\uparrow$\,2.68\%}} & 0.4712 & \textbf{\textcolor{blue}{$\downarrow$\,5.63\%}}& 0.8021 & \textbf{\textcolor{red}{$\uparrow$\,3.60\%}} & 0.5344 & \textbf{\textcolor{blue}{$\downarrow$\,4.76\%}} \\
 \midrule
 \multirow{4}{*}{DCN \cite{wang2017deep}} & Base & \rule[0.5ex]{0.6cm}{0.4pt}
 & 0.8271 & \rule[0.5ex]{0.6cm}{0.4pt} & 0.4991 & \rule[0.5ex]{0.6cm}{0.4pt}& 0.7746 & \rule[0.5ex]{0.6cm}{0.4pt}& 0.5605 &\rule[0.5ex]{0.6cm}{0.4pt} \\
 & KAR & GPT-3.5
 & 0.8350 & \textcolor{red!60}{$\uparrow$\,0.96\%} & 0.4918 & \textcolor{blue!60}{$\downarrow$\,1.46\%} & 0.7951 & \textcolor{red!60}{$\uparrow$\,2.65\%}& 0.5482 & \textcolor{blue!60}{$\downarrow$\,2.19\%} \\
 & $\mathcal{R}^2$ec & Qwen2.5-7B
 & 0.8431 & \textcolor{red!60}{$\uparrow$\,1.93\%} & 0.4885 & \textcolor{blue!60}{$\downarrow$\,2.12\%} & 0.7959 & \textcolor{red!60}{$\uparrow$\,2.75\%}& 0.5400 & \textcolor{blue!60}{$\downarrow$\,3.66\%}\\
 & $\mathcal{R}^4$ec & Qwen2.5-7B
 & 0.8476 & \textbf{\textcolor{red}{$\uparrow$\,2.48\%}} & 0.4754 & \textbf{\textcolor{blue}{$\downarrow$\,4.75\%}}& 0.8007 & \textbf{\textcolor{red}{$\uparrow$\,3.37\%}}& 0.5349 & \textbf{\textcolor{blue}{$\downarrow$\,4.57\%}}\\
 \midrule
 \multirow{4}{*}{DeepFM \cite{guo2017deepfm}} & Base & \rule[0.5ex]{0.6cm}{0.4pt}
 & 0.8276 & \rule[0.5ex]{0.6cm}{0.4pt} & 0.4986 & \rule[0.5ex]{0.6cm}{0.4pt}& 0.7740 & \rule[0.5ex]{0.6cm}{0.4pt}& 0.5616 & \rule[0.5ex]{0.6cm}{0.4pt}\\
 & KAR & GPT-3.5
 & 0.8370 & \textcolor{red!60}{$\uparrow$\,1.14\%} & 0.4858 & \textcolor{blue!60}{$\downarrow$\,2.56\%} & 0.7953 & \textcolor{red!60}{$\uparrow$\,2.73\%} & 0.5397 & \textcolor{blue!60}{$\downarrow$\,3.89\%}\\
 & $\mathcal{R}^2$ec & Qwen2.5-7B
 & 0.8454 & \textcolor{red!60}{$\uparrow$\,2.15\%} & 0.4779 & \textcolor{blue!60}{$\downarrow$\,4.15\%} & 0.7940 & \textcolor{red!60}{$\uparrow$\,2.82\%} & 0.5403 & \textcolor{blue!60}{$\downarrow$\,3.79\%}\\
 & $\mathcal{R}^4$ec & Qwen2.5-7B
 & 0.8483 & \textbf{\textcolor{red}{$\uparrow$\,2.50\%}} & 0.4704 & \textbf{\textcolor{blue}{$\downarrow$\,5.66\%}} & 0.7998 & \textbf{\textcolor{red}{$\uparrow$\,3.33\%}}& 0.5366 & \textbf{\textcolor{blue}{$\downarrow$\,4.45\%}}\\
\bottomrule
\end{tabular}
\caption{Experimental results for different CTR backbones on Amazon-Book and MovieLens-1M datasets. We report AUC and Logloss. "Rel. Impr." is the relative improvement rate of method against each base model.}
\label{tab:main results}
\end{table*}



The datasets used in this paper are described as follows: \textbf{Amazon-Book} is the "Book" category of the Amazon Review Dataset. We regard reviews with ratings greater than 5 with positive.  \textbf{MovieLens-1M} is a movie recommendation dataset with user-movie ratings ranging from 1 to 5. Samples with ratings greater than 3 are labeled as positive with the rest as negative.  \textbf{Industrial Dataset} is collected from a large-scale advertising platform with hundreds of millions of users. Samples are constructed through sampling from impression logs.  For academic datasets, we sort all interaction behaviors in chronological order and take the first 80\% as the training set and the remaining as the testing set.
Detailed statistics of the datasets are shown in Table \ref{tab:statistics}.



\subsubsection{Backbone Models}\label{backbone model introduction}
Because $\mathcal{R}^{4}$ec is a model-agnostic framework, various classic models can serve as the backbones.  Here, we choose 6 representative CTR models.  A brief introduction is provided below:  \textbf{DIEN} \cite{zhou2019deep} incorporates an interest evolution mechanism to capture the dynamic evolution of user interests over time.  \textbf{GRU4Rec} \cite{hidasi2015session} employs Gated Recurrent Units (GRU) combined with a ranking-based loss function to effectively model user sequences for recommendation systems.  \textbf{AutoInt} \cite{song2019autoint} employs a self-attentive network enhanced with residual connections to model feature interactions. 
\textbf{FiGNN} \cite{li2019fi} designs a novel model feature interaction graph network to utilize the strong representative of graphs.  \textbf{DCN} \cite{wang2017deep} leverages a cross-network architecture to capture the bounded-degree feature interactions. \textbf{DeepFM} \cite{guo2017deepfm} adopts factorization machine to capture low-order and high-order feature interactions.

\subsubsection{Evaluation Metrics}
We utilize widely-used AUC (Area under the ROC curve) and LogLoss (binary cross-entropy loss) as evaluation metrics following previous studies \cite{xi2024towards,guo2017deepfm,song2019autoint}. \textbf{A higher AUC value or a lower Logloss value, even by a small margin (0.001 for example), can be viewed as a significant improvement}, as indicated by prior research \cite{xi2024towards}.

\subsubsection{Baselines}
We compare $\mathcal{R}^{4}$ec with the following settings: \textbf{Base}: we simply conduct experiments on conventional recommendation backbones that are introduced in Sec. \ref{backbone model introduction} without knowledge from LLMs.  \textbf{KAR}: Kar \cite{xi2024towards} acquires knowledge about users and items via the chain-of-thought technique, these knowledge will be employed as augment features for recommendation tasks. \textbf{$\boldsymbol{\mathcal{R}^2}$ec}: we employ only the obtained $\mathcal{D}_{reason}$ for training a single actor, i.e., without the reflection and refinement mechanisms. 


\subsubsection{Implementation Details}
For dataset construction, we utilize API of a widely-used LLM \textit{gpt-4o}. For the construction of the reasoning, reflection, and refinement dataset, we construct data from 40\% of the users and items, and for each user or item, we generate one supervision sample.  Additionally, we perform two separate inferences for each sample.  We employ Qwen-2.5 7B \cite{yang2024qwen2} as our default actor model and reflection model. During training these LLMs, we set LoRA rank $r$ as 8, LoRA alpha $\alpha$ as 16, and LoRA dropout as 0.05. The LoRA update matrices are applied on all linear layers. We fine-tune the LLMs for 3 epochs.  During inference, we employ an iterative refinement strategy with the number of iterations set to 1 by default.
Our implementations for conventional recommendation backbones follow \cite{xi2024towards}. For knowledge encoder, we use BGE-M3 \cite{chen2024bge} by default.  Please refer our implementation code for more details, including prompt templates and so on.

 

            

\subsection{Experimental Results}

We implement $\mathcal{R}^{4}$ec, $\mathcal{R}^{2}$ec, and KAR \cite{xi2024towards} upon 6 representative CTR models.  The experimental results are showcased in Table \ref{tab:main results}, which includes the AUC and its relative improvements, as well as the LogLoss and its relative reductions on both Amazon-Book and MovieLens-1M datasets.  From these results, we can draw the following observations:

\begin{itemize}[leftmargin=*]
    \item Extracting user preference knowledge and item factual knowledge from LLMs significantly enhances the performance of recommendation systems. $\mathcal{R}^{4}$ec, $\mathcal{R}^{2}$ec, and KAR, designed to extract knowledge about users and items from LLMs, have facilitated the incorporation of this knowledge as sideinfo into downstream recommendation backbones.  This integration has led to improvements in AUC and reductions in LogLoss across both the Amazon-Book and MovieLens-1M.  
    \item Despite $\mathcal{R}^{4}$ec using the Qwen2.5-7b model and KAR employing GPT-3.5, $\mathcal{R}^{4}$ec achieves more pronounced improvements in AUC and reductions in LogLoss compared to KAR.  Specifically, $\mathcal{R}^{4}$ec achieves a 1.36\% greater relative improvement in AUC and a 3.29\% greater relative reduction in LogLoss than KAR.  We speculate that the more pronounced improvements with $\mathcal{R}^{4}$ec compared to KAR stem from KAR's direct prompting of GPT-3.5 for knowledge acquisition, which might lead to hallucinations. In contrast, our $\mathcal{R}^{4}$ec can continuously identify and resolve issues, thereby significantly enhancing the reliability of the extraction of user preference knowledge and item factual knowledge.
    \item Comparing the $\mathcal{R}^{4}$ec  with $\mathcal{R}^{2}$ec reveals that the user preference knowledge and item factual knowledge refined through our iterative reflection and refinement mechanism can lead to more significant improvements in downstream recommendation backbones.  Specifically, on the Amazon-Book dataset, the $\mathcal{R}^{4}$ec outperformed $\mathcal{R}^{2}$ec, showing a 0.60\% higher relative increase in AUC and a 1.93\% greater relative reduction in LogLoss, averaged across six backbone models.   Similarly, on the MovieLens-1M dataset, the $\mathcal{R}^{4}$ec demonstrated a 0.62\% greater relative improvement in AUC and a 0.68\% larger relative reduction in LogLoss compared to $\mathcal{R}^{2}$ec. These results validate the superiority of System-2 thinking over System-1 thinking, demonstrating significant potential in exploring System-2 thinking in recommendation systems.
\end{itemize}

\subsection{Online Experimental} 

\begin{table}[h]
\centering
\renewcommand{\arraystretch}{0.85}
\begin{tabular}{@{}lccr@{}}
\toprule
Method &  Setting & Revenue & CVR \\ 
\midrule
\multirow{2}{*}{$\mathcal{R}^4$ec} & all & \textcolor{red}{$\uparrow$\,2.2\%} & \textcolor{red}{$\uparrow$\,1.6\%} \\
       & long-tail &  \textcolor{red}{$\uparrow$\,4.1\%} & \textcolor{red}{$\uparrow$\,3.2\%} \\
\bottomrule
\end{tabular}
\caption{Results on online advertising platform.}
\label{Results on industrial dataset}
\end{table}

\begin{figure*}[t]  
\centering  
\includegraphics[width=\textwidth]{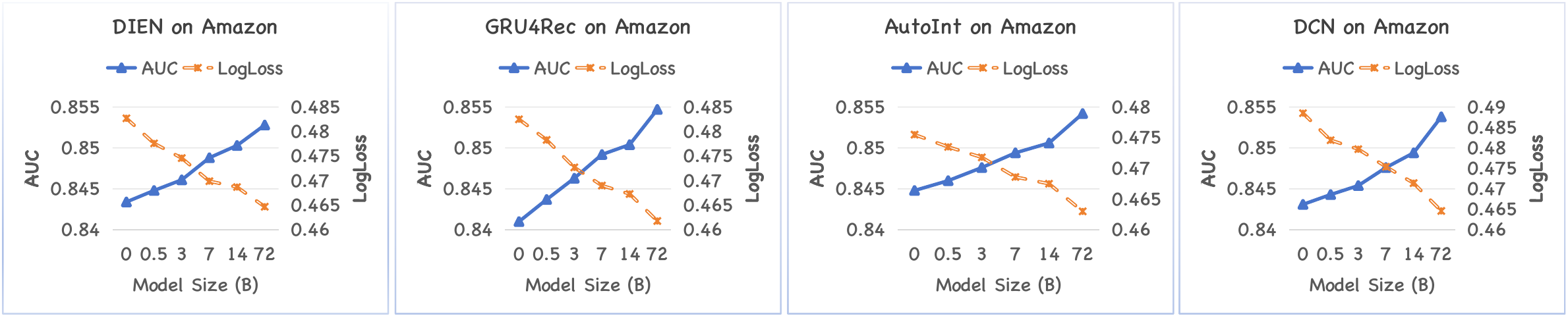} 
\caption{We use the Qwen-2.5 7B model as the actor and the Qwen-2.5 models with sizes 0.5B, 3B, 7B, 14B, and 72B as the reflection models.  The AUC and LogLoss performance on the Amazon-Book dataset are shown.}  
\label{fig:scaling properties of reflection models}  
\end{figure*}

\subsubsection{Experimental Setup}
To validate the effectiveness of $\mathcal{R}^4$ec in real-world scenarios, we conduct online A/B experiments on a large-scale advertising platform.  The traffic of the whole app is split into ten buckets uniformly.  20\% of traffic is assigned to the online baseline while anther 10\% is assigned to $\mathcal{R}^4$ec.  As revealed in Tab. \ref{tab:statistics}, our advertising platform serves over 400 million users, and the results collected from 10\% of traffic for several weeks are convincing.  We randomly sample one million samples for user and item reasoning, reflection, and refinement datasets construction, respectively.  For the item factual knowledge extraction, we perform LLM inference across all items. In contrast, extracting user preference knowledge, which involves processing hundreds of millions of data points, results in high inference costs. Therefore, we tailor our inference strategies to the activity levels of users on the advertising platform. For active users, we conduct full inference using the LLM. For less active users, since the item knowledge has already been fully inferred, we approximate the inference results of the item knowledge in the user's historical behavior list as the inference results for their preference.

\subsubsection{Experimental Results}
As indicated in Tab. \ref{Results on industrial dataset}, in a 14-day online A/B test, our method exhibits a 2.2\% increase of revenue and 1.6\% improvement of the conversion rate compared with the baseline, resulting in significant business benefits.  Furthermore, we utilize data with limited interactions to verify the effectiveness of cold starts.  Experiment results in Tab. \ref{Results on industrial dataset} demonstrate that $\mathcal{R}^4$ec can achieve a 4.1\% increase of revenue and 3.2\% improvement of conversion rate on long-tail data. This indicates that $\mathcal{R}^4$ec can be successfully implemented in industrial settings and improve recommendation experience for real-world users.

\subsection{Ablation Study}

\begin{table}[htbp]
\centering
\renewcommand{\arraystretch}{0.9}
\resizebox{\columnwidth}{!}{
\begin{tabular}{c|c|cc|cc}
\toprule
\multirow{2}{*}{Backbone} & \multirow{2}{*}{Encoder} & \multicolumn{2}{c}{Amazon-Book} & \multicolumn{2}{c}{MovieLens-1M} \\
& & AUC & LogLoss & AUC & LogLoss\\
\midrule
\multirow{4}{*}{AutoInt}  & Base & 0.8261 & 0.5007 &  0.7736 & 0.5618 \\
& Bert & 0.8449 & 0.4768 & 0.7934 & 0.5412 \\
& Longformer & 0.8462 & 0.4760 & 0.7953 & 0.5395 \\
& BGE-M3 & 0.8494 & 0.4686 & 0.8008 & 0.5347 \\
\bottomrule
\end{tabular}
}
\caption{Experimental results of different knowledge encoders on Amazon-Book and MovieLens-1M.  We report AUC and LogLoss.}
\label{tab:different backbone}
\end{table}

\subsubsection{Effect of Different Knowledge Encoders}
In this section, to investigate the impact of different knowledge encoders, we compare the AUC and LogLoss results of AutoInt on Amazon-Book and MovieLens-1M using BERT \cite{devlin2018bert}, Longformer \cite{beltagy2020longformer}, and BGE-M3 \cite{chen2024bge}as knowledge encoders. The experimental results are presented in Tab. \ref{tab:different backbone}.

From Tab. \ref{tab:different backbone}, we observe the following: (1) Regardless of the knowledge encoder used, both AUC significantly increases and LogLoss significantly decreases, demonstrating that extracting user preference and item factual knowledge can improve recommendation system performance. (2) BGE-M3 yields the most significant improvements, highlighting its superiority as knowledge encoder for recommendation system.

\begin{figure}[t]  
\centering  
\includegraphics[width=0.45\textwidth]{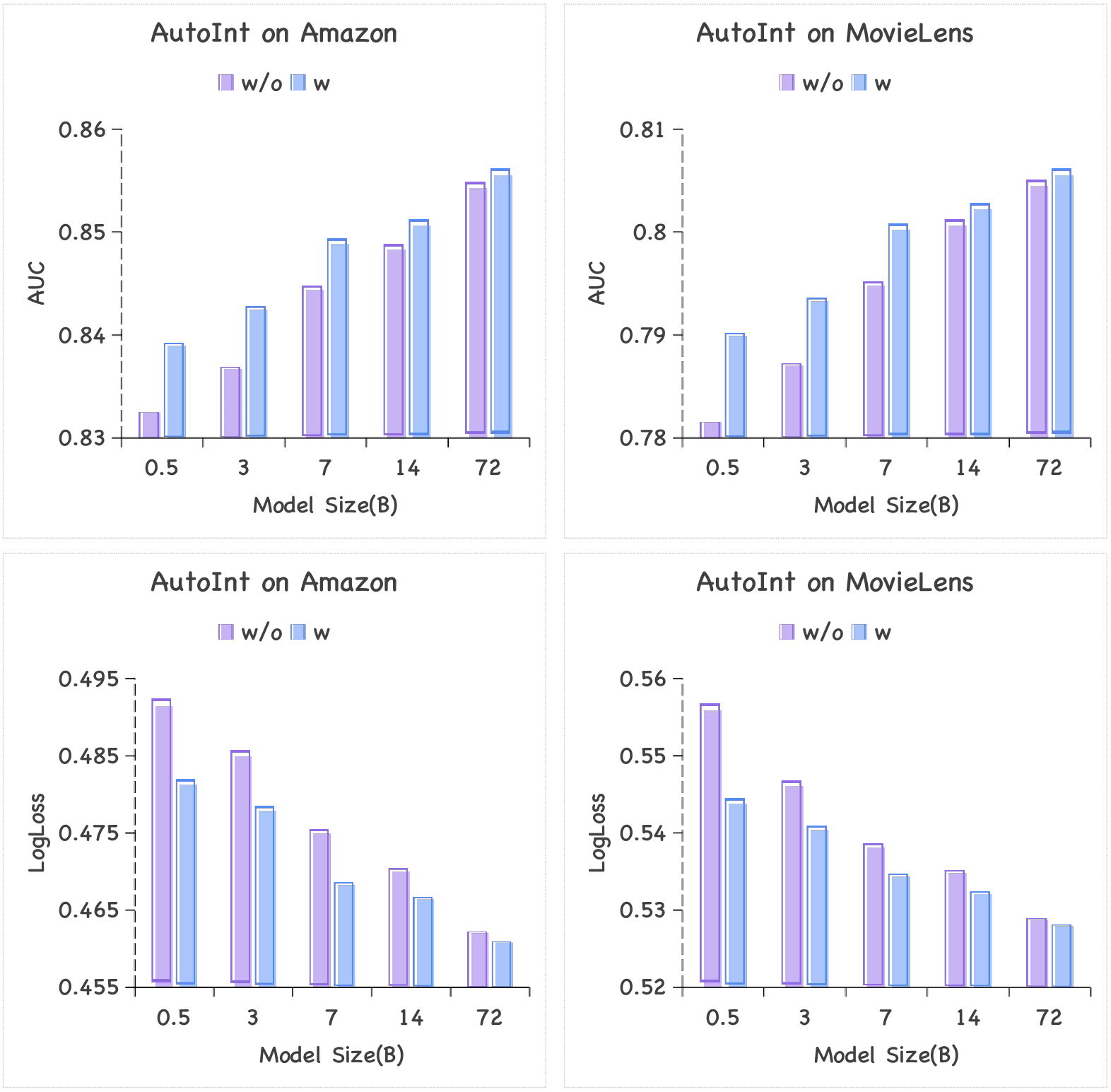} 
\caption{We use the Qwen-2.5 7B model as the reflection model and Qwen-2.5 models with sizes 0.5B, 3B, 7B, 14B, and 72B as the actor models.Here, 'w/o' and 'w/' denotes 'without a reflection model' and a 'with a reflection model', respectively.}  
\label{fig:scaling properties of actor models}  
\end{figure}

\subsubsection{Scaling Properties of Reflection Model}\label{Scaling Properties of Reflection Model}
To investigate the scaling law of the reflection model, namely, whether the performance improves as the size of the reflection model increases while the size of the actor model remains fixed. We conduct experiments on the Qwen-2.5 series models of varying sizes. The Qwen-2.5 7B is utilized as the actor model, paired with reflection models of sizes 0.5B, 3B, 7B, 14B, and 72B.  We employ DIEN, GRU4Rec, AutoInt, and DCN as recommendation backbones. The trends of AUC and LogLoss across different sizes of the reflection models on Amazon-Book are depicted in Fig. \ref{fig:scaling properties of reflection models}.

We find that as the size of the reflection model increases from 0.5B to 72B,  there is a consistent improvement in AUC alongside a notable reduction in LogLoss. Specifically, as the size of the reflection model is increased from 0.5B to 72B, the average AUC on the Amazon dataset improves by a relative 1.3\%, accompanied by a 3.8\% relative reduction in LogLoss across four recommendation backbones. This trend underscores the scaling properties of reflection model, indicating that a larger reflection model can provide better feedback, thereby yielding knowledge that is more suitable for downstream recommendation models. 


\subsubsection{Scaling Properties of Actor Model}

In this section, we further explore the scaling properties of actor models. To this end, we use the Qwen-2.5 7B model \cite{yang2024qwen2} as the reflection model and Qwen-2.5 models with sizes 0.5B, 3B, 7B, 14B, and 72B as the actor models. Fig. \ref{fig:scaling properties of actor models} presents the AUC and LogLoss performance on Amazon-Book and MovieLens. We conduct experiments on AutoInt.

From Fig. \ref{fig:scaling properties of actor models}, we observe the following: (1) Regardless of the actor model's scale, the inclusion of the 7B reflection model leads to consistent performance improvements. This result suggests that the 7B reflection model can consistently provide effective supervision, even when the actor model size reaches 72B. This indicates that even smaller reflection models can enhance larger actor models to a certain extent. (2) As the size of the actor model increases, the performance improvement from the reflection model tends to diminish. 


\begin{figure}[t]  
\centering  
\includegraphics[width=0.49\textwidth]{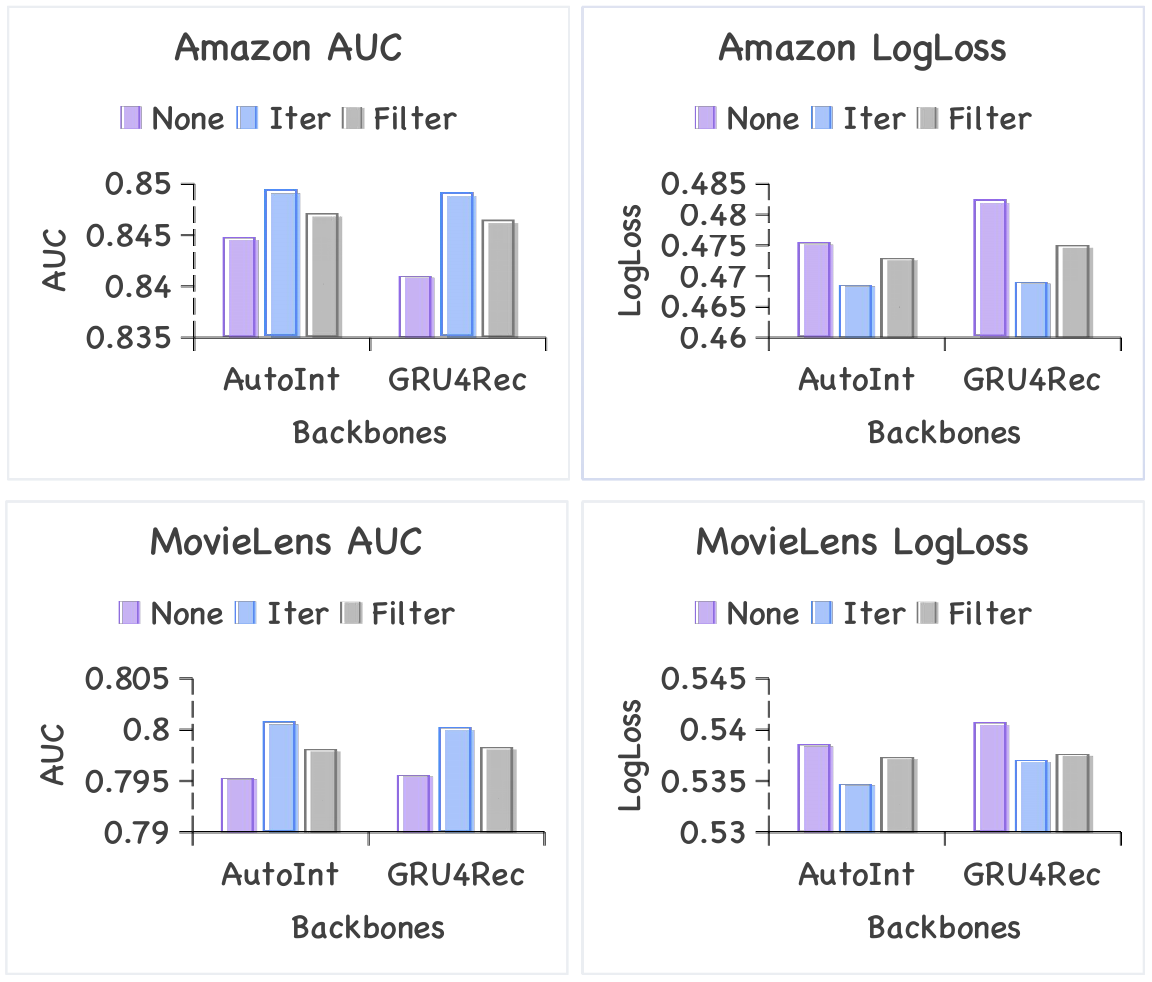} 
\caption{We compare the performance of AutoInt and GRU4Rec under the "Iterative Refinement" (Iter) and "Reflection as a Filter" (Filter) inference strategies.}  
\label{fig:inference strategy}  
\end{figure}

\subsubsection{Effect of Different Inference Strategies}
In Sec. \ref{Inference Strategy}, we introduce two inference strategies "Iterative Refinement" and "Reflection as a Filter". For brevity, we refer to them as "Iter" and "Filter" respectively.  In this section, we compare the performance of these two strategies. For "Filter" strategy, we perform three rounds of inference with the actor model, and the resulting knowledge is subsequently filtered through the reflection model. Fig. \ref{fig:inference strategy} presents the performance comparison of AutoInt and GRU4Rec on the Amazon and MovieLens datasets.

From Fig. \ref{fig:inference strategy}, we can draw the following conclusions: (1) Both inference strategies lead to significant performance improvements, validating that the reflection mechanism helps acquire more reasonable and effective knowledge. (2) The "Iter" strategy consistently outperforms the "Filter" strategy, suggesting that using reflection alone is insufficient. Instead, it should be combined with refinement to correct unreasonable knowledge in recommendation system.

\subsubsection{Effect of Iterative Refinement Steps}
\begin{figure}[t]  
\centering  
\includegraphics[width=0.49\textwidth]{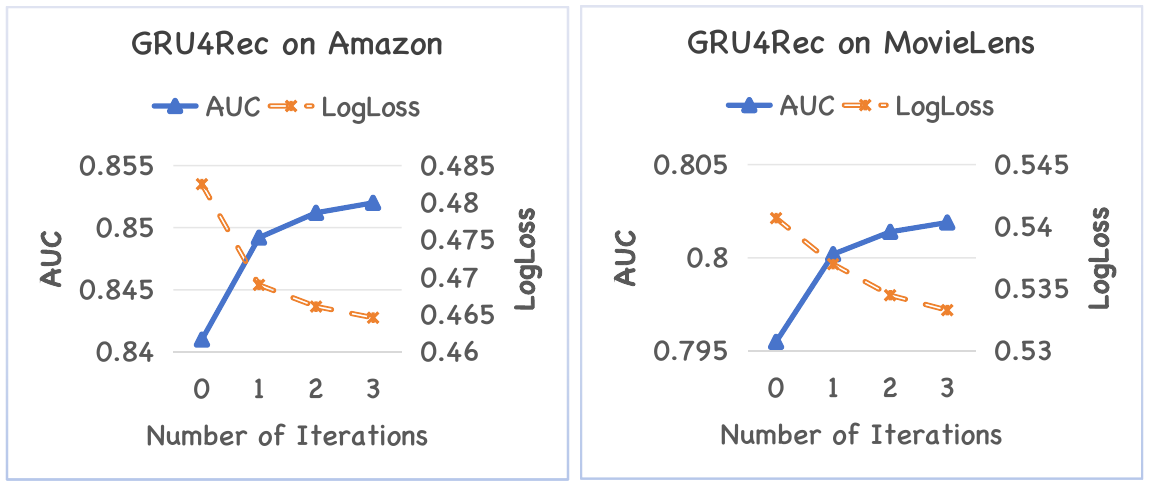} 
\caption{GRU4Rec's AUC and LogLoss performance on Amazon-Book and MovieLens with respect to the number of iterative refinement steps. We report AUC and LogLoss.}  
\label{fig:number-of-iters}  
\end{figure}

Since our iterative refinement inference strategy allows the reflection and actor models to continuously reflect and refine, the number of iterations becomes a critical hyperparameter. In this section, we investigate the impact of the number of iterative refinement steps on the performance of downstream recommendation models. Fig. \ref{fig:number-of-iters} shows the performance of GRU4Rec on Amazon-Book and MovieLens-1M.  

From Fig. \ref{fig:number-of-iters}, we observe that (1) by scaling up inference-time computation, i.e., increasing the number of iterative refinement steps, continuous improvements in AUC and reductions in LogLoss are achieved. This suggests that with each successive refinement, the acquired knowledge becomes more useful and rational. (2) As the number of iterations increases, the rate of improvement in AUC increases at a diminishing rate.  We hypothesize that after multiple refinements, the reflection model's capacity becomes a bottleneck, , leading to the majority of the knowledge generated by the actor model being deemed reasonable by the reflection model.  Finally, considering the scale of users and items in the recommendation system, as well as time and inference cost constraints, we adopt a single refinement step as the default method.

\section{Conclusion}
In this paper, we propose $\mathcal{R}^4$ec,  a reasoning, reflection, and refinement framework that explores System-2 thinking within recommendation systems. Specifically, we introduce two models: the actor model, responsible for reasoning, and the reflection model, which evaluates the reasonableness of the actor's responses and provides feedback. This feedback encourages the actor model to refine its responses. Through this iterative process of reflection and refinement, we facilitate a slow and deliberate thinking process akin to System-2, ensuring more accurate acquisition of user preferences and factual item information. Ultimately, the backbone recommendation model integrates the refined knowledge from LLMs with original categorical features to improve recommendation performance.  We demonstrate the effectiveness of $\mathcal{R}^4$ec through substantial improvements across Amazon-Book and MovieLens-1M.  Additionally, our results validate the scaling properties of the actor model and reflection model.  We hope that our work will inspire further research into advancing System-2 thinking in recommendation systems.

\bibliographystyle{ACM-Reference-Format}
\bibliography{sample-base}


\begin{thebibliography}{59}


\ifx \showCODEN    \undefined \def \showCODEN     #1{\unskip}     \fi
\ifx \showISBNx    \undefined \def \showISBNx     #1{\unskip}     \fi
\ifx \showISBNxiii \undefined \def \showISBNxiii  #1{\unskip}     \fi
\ifx \showISSN     \undefined \def \showISSN      #1{\unskip}     \fi
\ifx \showLCCN     \undefined \def \showLCCN      #1{\unskip}     \fi
\ifx \shownote     \undefined \def \shownote      #1{#1}          \fi
\ifx \showarticletitle \undefined \def \showarticletitle #1{#1}   \fi
\ifx \showURL      \undefined \def \showURL       {\relax}        \fi
\providecommand\bibfield[2]{#2}
\providecommand\bibinfo[2]{#2}
\providecommand\natexlab[1]{#1}
\providecommand\showeprint[2][]{arXiv:#2}

\bibitem[Achiam et~al\mbox{.}(2023)]%
        {achiam2023gpt}
\bibfield{author}{\bibinfo{person}{Josh Achiam}, \bibinfo{person}{Steven Adler}, \bibinfo{person}{Sandhini Agarwal}, \bibinfo{person}{Lama Ahmad}, \bibinfo{person}{Ilge Akkaya}, \bibinfo{person}{Florencia~Leoni Aleman}, \bibinfo{person}{Diogo Almeida}, \bibinfo{person}{Janko Altenschmidt}, \bibinfo{person}{Sam Altman}, \bibinfo{person}{Shyamal Anadkat}, {et~al\mbox{.}}} \bibinfo{year}{2023}\natexlab{}.
\newblock \showarticletitle{Gpt-4 technical report}.
\newblock \bibinfo{journal}{\emph{arXiv preprint arXiv:2303.08774}} (\bibinfo{year}{2023}).
\newblock


\bibitem[Bai et~al\mbox{.}(2023)]%
        {bai2023qwen}
\bibfield{author}{\bibinfo{person}{Jinze Bai}, \bibinfo{person}{Shuai Bai}, \bibinfo{person}{Yunfei Chu}, \bibinfo{person}{Zeyu Cui}, \bibinfo{person}{Kai Dang}, \bibinfo{person}{Xiaodong Deng}, \bibinfo{person}{Yang Fan}, \bibinfo{person}{Wenbin Ge}, \bibinfo{person}{Yu Han}, \bibinfo{person}{Fei Huang}, {et~al\mbox{.}}} \bibinfo{year}{2023}\natexlab{}.
\newblock \showarticletitle{Qwen technical report}.
\newblock \bibinfo{journal}{\emph{arXiv preprint arXiv:2309.16609}} (\bibinfo{year}{2023}).
\newblock


\bibitem[Bai et~al\mbox{.}(2024)]%
        {bai2024finetuning}
\bibfield{author}{\bibinfo{person}{Zhuoxi Bai}, \bibinfo{person}{Ning Wu}, \bibinfo{person}{Fengyu Cai}, \bibinfo{person}{Xinyi Zhu}, {and} \bibinfo{person}{Yun Xiong}.} \bibinfo{year}{2024}\natexlab{}.
\newblock \showarticletitle{Finetuning Large Language Model for Personalized Ranking}.
\newblock \bibinfo{journal}{\emph{arXiv preprint arXiv:2405.16127}} (\bibinfo{year}{2024}).
\newblock


\bibitem[Bao et~al\mbox{.}(2023)]%
        {bao2023tallrec}
\bibfield{author}{\bibinfo{person}{Keqin Bao}, \bibinfo{person}{Jizhi Zhang}, \bibinfo{person}{Yang Zhang}, \bibinfo{person}{Wenjie Wang}, \bibinfo{person}{Fuli Feng}, {and} \bibinfo{person}{Xiangnan He}.} \bibinfo{year}{2023}\natexlab{}.
\newblock \showarticletitle{Tallrec: An effective and efficient tuning framework to align large language model with recommendation}. In \bibinfo{booktitle}{\emph{Proceedings of the 17th ACM Conference on Recommender Systems}}. \bibinfo{pages}{1007--1014}.
\newblock


\bibitem[Beltagy et~al\mbox{.}(2020)]%
        {beltagy2020longformer}
\bibfield{author}{\bibinfo{person}{Iz Beltagy}, \bibinfo{person}{Matthew~E Peters}, {and} \bibinfo{person}{Arman Cohan}.} \bibinfo{year}{2020}\natexlab{}.
\newblock \showarticletitle{Longformer: The long-document transformer}.
\newblock \bibinfo{journal}{\emph{arXiv preprint arXiv:2004.05150}} (\bibinfo{year}{2020}).
\newblock


\bibitem[Chen et~al\mbox{.}(2024)]%
        {chen2024bge}
\bibfield{author}{\bibinfo{person}{Jianlv Chen}, \bibinfo{person}{Shitao Xiao}, \bibinfo{person}{Peitian Zhang}, \bibinfo{person}{Kun Luo}, \bibinfo{person}{Defu Lian}, {and} \bibinfo{person}{Zheng Liu}.} \bibinfo{year}{2024}\natexlab{}.
\newblock \showarticletitle{Bge m3-embedding: Multi-lingual, multi-functionality, multi-granularity text embeddings through self-knowledge distillation}.
\newblock \bibinfo{journal}{\emph{arXiv preprint arXiv:2402.03216}} (\bibinfo{year}{2024}).
\newblock


\bibitem[Chen et~al\mbox{.}(2023)]%
        {chen2023teaching}
\bibfield{author}{\bibinfo{person}{Xinyun Chen}, \bibinfo{person}{Maxwell Lin}, \bibinfo{person}{Nathanael Sch{\"a}rli}, {and} \bibinfo{person}{Denny Zhou}.} \bibinfo{year}{2023}\natexlab{}.
\newblock \showarticletitle{Teaching large language models to self-debug}.
\newblock \bibinfo{journal}{\emph{arXiv preprint arXiv:2304.05128}} (\bibinfo{year}{2023}).
\newblock


\bibitem[Dai et~al\mbox{.}(2023)]%
        {dai2023uncovering}
\bibfield{author}{\bibinfo{person}{Sunhao Dai}, \bibinfo{person}{Ninglu Shao}, \bibinfo{person}{Haiyuan Zhao}, \bibinfo{person}{Weijie Yu}, \bibinfo{person}{Zihua Si}, \bibinfo{person}{Chen Xu}, \bibinfo{person}{Zhongxiang Sun}, \bibinfo{person}{Xiao Zhang}, {and} \bibinfo{person}{Jun Xu}.} \bibinfo{year}{2023}\natexlab{}.
\newblock \showarticletitle{Uncovering chatgpt’s capabilities in recommender systems}. In \bibinfo{booktitle}{\emph{Proceedings of the 17th ACM Conference on Recommender Systems}}. \bibinfo{pages}{1126--1132}.
\newblock


\bibitem[Devlin(2018)]%
        {devlin2018bert}
\bibfield{author}{\bibinfo{person}{Jacob Devlin}.} \bibinfo{year}{2018}\natexlab{}.
\newblock \showarticletitle{Bert: Pre-training of deep bidirectional transformers for language understanding}.
\newblock \bibinfo{journal}{\emph{arXiv preprint arXiv:1810.04805}} (\bibinfo{year}{2018}).
\newblock


\bibitem[Du et~al\mbox{.}(2024)]%
        {du2024enhancing}
\bibfield{author}{\bibinfo{person}{Yingpeng Du}, \bibinfo{person}{Di Luo}, \bibinfo{person}{Rui Yan}, \bibinfo{person}{Xiaopei Wang}, \bibinfo{person}{Hongzhi Liu}, \bibinfo{person}{Hengshu Zhu}, \bibinfo{person}{Yang Song}, {and} \bibinfo{person}{Jie Zhang}.} \bibinfo{year}{2024}\natexlab{}.
\newblock \showarticletitle{Enhancing job recommendation through llm-based generative adversarial networks}. In \bibinfo{booktitle}{\emph{Proceedings of the AAAI Conference on Artificial Intelligence}}, Vol.~\bibinfo{volume}{38}. \bibinfo{pages}{8363--8371}.
\newblock


\bibitem[Gao et~al\mbox{.}(2023)]%
        {gao2023chat}
\bibfield{author}{\bibinfo{person}{Yunfan Gao}, \bibinfo{person}{Tao Sheng}, \bibinfo{person}{Youlin Xiang}, \bibinfo{person}{Yun Xiong}, \bibinfo{person}{Haofen Wang}, {and} \bibinfo{person}{Jiawei Zhang}.} \bibinfo{year}{2023}\natexlab{}.
\newblock \showarticletitle{Chat-rec: Towards interactive and explainable llms-augmented recommender system}.
\newblock \bibinfo{journal}{\emph{arXiv preprint arXiv:2303.14524}} (\bibinfo{year}{2023}).
\newblock


\bibitem[Geng et~al\mbox{.}(2022)]%
        {geng2022recommendation}
\bibfield{author}{\bibinfo{person}{Shijie Geng}, \bibinfo{person}{Shuchang Liu}, \bibinfo{person}{Zuohui Fu}, \bibinfo{person}{Yingqiang Ge}, {and} \bibinfo{person}{Yongfeng Zhang}.} \bibinfo{year}{2022}\natexlab{}.
\newblock \showarticletitle{Recommendation as language processing (rlp): A unified pretrain, personalized prompt \& predict paradigm (p5)}. In \bibinfo{booktitle}{\emph{Proceedings of the 16th ACM Conference on Recommender Systems}}. \bibinfo{pages}{299--315}.
\newblock


\bibitem[Gou et~al\mbox{.}(2023)]%
        {gou2023critic}
\bibfield{author}{\bibinfo{person}{Zhibin Gou}, \bibinfo{person}{Zhihong Shao}, \bibinfo{person}{Yeyun Gong}, \bibinfo{person}{Yelong Shen}, \bibinfo{person}{Yujiu Yang}, \bibinfo{person}{Nan Duan}, {and} \bibinfo{person}{Weizhu Chen}.} \bibinfo{year}{2023}\natexlab{}.
\newblock \showarticletitle{Critic: Large language models can self-correct with tool-interactive critiquing}.
\newblock \bibinfo{journal}{\emph{arXiv preprint arXiv:2305.11738}} (\bibinfo{year}{2023}).
\newblock


\bibitem[Gu et~al\mbox{.}(2025)]%
        {gu2025mathcal}
\bibfield{author}{\bibinfo{person}{Hao Gu}, \bibinfo{person}{Jiangyan Yi}, \bibinfo{person}{Chenglong Wang}, \bibinfo{person}{Jianhua Tao}, \bibinfo{person}{Zheng Lian}, \bibinfo{person}{Jiayi He}, \bibinfo{person}{Yong Ren}, \bibinfo{person}{Yujie Chen}, {and} \bibinfo{person}{Zhengqi Wen}.} \bibinfo{year}{2025}\natexlab{}.
\newblock \showarticletitle{ALLM4ADD: Unlocking the Capabilities of Audio Large Language Models for Audio Deepfake Detection}.
\newblock \bibinfo{journal}{\emph{arXiv preprint arXiv:2505.11079}} (\bibinfo{year}{2025}).
\newblock


\bibitem[Guo et~al\mbox{.}(2017)]%
        {guo2017deepfm}
\bibfield{author}{\bibinfo{person}{Huifeng Guo}, \bibinfo{person}{Ruiming Tang}, \bibinfo{person}{Yunming Ye}, \bibinfo{person}{Zhenguo Li}, {and} \bibinfo{person}{Xiuqiang He}.} \bibinfo{year}{2017}\natexlab{}.
\newblock \showarticletitle{DeepFM: a factorization-machine based neural network for CTR prediction}.
\newblock \bibinfo{journal}{\emph{arXiv preprint arXiv:1703.04247}} (\bibinfo{year}{2017}).
\newblock


\bibitem[He et~al\mbox{.}(2020)]%
        {he2020lightgcn}
\bibfield{author}{\bibinfo{person}{Xiangnan He}, \bibinfo{person}{Kuan Deng}, \bibinfo{person}{Xiang Wang}, \bibinfo{person}{Yan Li}, \bibinfo{person}{Yongdong Zhang}, {and} \bibinfo{person}{Meng Wang}.} \bibinfo{year}{2020}\natexlab{}.
\newblock \showarticletitle{Lightgcn: Simplifying and powering graph convolution network for recommendation}. In \bibinfo{booktitle}{\emph{Proceedings of the 43rd International ACM SIGIR conference on research and development in Information Retrieval}}. \bibinfo{pages}{639--648}.
\newblock


\bibitem[Hegel(1991)]%
        {hegel1991encyclopaedia}
\bibfield{author}{\bibinfo{person}{Georg Wilhelm~Friedrich Hegel}.} \bibinfo{year}{1991}\natexlab{}.
\newblock \bibinfo{booktitle}{\emph{The Encyclopaedia Logic, with the Zus tze: Part I of the Encyclopaedia of Philosophical Sciences with the Zus{\"a}tze}}. Vol.~\bibinfo{volume}{1}.
\newblock \bibinfo{publisher}{Hackett Publishing}.
\newblock


\bibitem[Hidasi(2015)]%
        {hidasi2015session}
\bibfield{author}{\bibinfo{person}{B Hidasi}.} \bibinfo{year}{2015}\natexlab{}.
\newblock \showarticletitle{Session-based Recommendations with Recurrent Neural Networks}.
\newblock \bibinfo{journal}{\emph{arXiv preprint arXiv:1511.06939}} (\bibinfo{year}{2015}).
\newblock


\bibitem[Hu et~al\mbox{.}(2021)]%
        {hu2021lora}
\bibfield{author}{\bibinfo{person}{Edward~J Hu}, \bibinfo{person}{Yelong Shen}, \bibinfo{person}{Phillip Wallis}, \bibinfo{person}{Zeyuan Allen-Zhu}, \bibinfo{person}{Yuanzhi Li}, \bibinfo{person}{Shean Wang}, \bibinfo{person}{Lu Wang}, {and} \bibinfo{person}{Weizhu Chen}.} \bibinfo{year}{2021}\natexlab{}.
\newblock \showarticletitle{Lora: Low-rank adaptation of large language models}.
\newblock \bibinfo{journal}{\emph{arXiv preprint arXiv:2106.09685}} (\bibinfo{year}{2021}).
\newblock


\bibitem[Ji et~al\mbox{.}(2025)]%
        {ji2025test}
\bibfield{author}{\bibinfo{person}{Yixin Ji}, \bibinfo{person}{Juntao Li}, \bibinfo{person}{Hai Ye}, \bibinfo{person}{Kaixin Wu}, \bibinfo{person}{Jia Xu}, \bibinfo{person}{Linjian Mo}, {and} \bibinfo{person}{Min Zhang}.} \bibinfo{year}{2025}\natexlab{}.
\newblock \showarticletitle{Test-time Computing: from System-1 Thinking to System-2 Thinking}.
\newblock \bibinfo{journal}{\emph{arXiv preprint arXiv:2501.02497}} (\bibinfo{year}{2025}).
\newblock


\bibitem[Jiang et~al\mbox{.}(2024)]%
        {jiang2024training}
\bibfield{author}{\bibinfo{person}{Nan Jiang}, \bibinfo{person}{Xiaopeng Li}, \bibinfo{person}{Shiqi Wang}, \bibinfo{person}{Qiang Zhou}, \bibinfo{person}{Soneya~Binta Hossain}, \bibinfo{person}{Baishakhi Ray}, \bibinfo{person}{Varun Kumar}, \bibinfo{person}{Xiaofei Ma}, {and} \bibinfo{person}{Anoop Deoras}.} \bibinfo{year}{2024}\natexlab{}.
\newblock \showarticletitle{Training LLMs to Better Self-Debug and Explain Code}.
\newblock \bibinfo{journal}{\emph{arXiv preprint arXiv:2405.18649}} (\bibinfo{year}{2024}).
\newblock


\bibitem[Kahneman(2011)]%
        {kahneman2011thinking}
\bibfield{author}{\bibinfo{person}{Daniel Kahneman}.} \bibinfo{year}{2011}\natexlab{}.
\newblock \showarticletitle{Thinking, fast and slow}.
\newblock \bibinfo{journal}{\emph{Farrar, Straus and Giroux}} (\bibinfo{year}{2011}).
\newblock


\bibitem[Ke et~al\mbox{.}(2023)]%
        {ke2023critiquellm}
\bibfield{author}{\bibinfo{person}{Pei Ke}, \bibinfo{person}{Bosi Wen}, \bibinfo{person}{Zhuoer Feng}, \bibinfo{person}{Xiao Liu}, \bibinfo{person}{Xuanyu Lei}, \bibinfo{person}{Jiale Cheng}, \bibinfo{person}{Shengyuan Wang}, \bibinfo{person}{Aohan Zeng}, \bibinfo{person}{Yuxiao Dong}, \bibinfo{person}{Hongning Wang}, {et~al\mbox{.}}} \bibinfo{year}{2023}\natexlab{}.
\newblock \showarticletitle{Critiquellm: Scaling llm-as-critic for effective and explainable evaluation of large language model generation}.
\newblock \bibinfo{journal}{\emph{arXiv preprint arXiv:2311.18702}} (\bibinfo{year}{2023}).
\newblock


\bibitem[Kim et~al\mbox{.}(2024)]%
        {kim2024language}
\bibfield{author}{\bibinfo{person}{Geunwoo Kim}, \bibinfo{person}{Pierre Baldi}, {and} \bibinfo{person}{Stephen McAleer}.} \bibinfo{year}{2024}\natexlab{}.
\newblock \showarticletitle{Language models can solve computer tasks}.
\newblock \bibinfo{journal}{\emph{Advances in Neural Information Processing Systems}}  \bibinfo{volume}{36} (\bibinfo{year}{2024}).
\newblock


\bibitem[Koren et~al\mbox{.}(2009)]%
        {koren2009matrix}
\bibfield{author}{\bibinfo{person}{Yehuda Koren}, \bibinfo{person}{Robert Bell}, {and} \bibinfo{person}{Chris Volinsky}.} \bibinfo{year}{2009}\natexlab{}.
\newblock \showarticletitle{Matrix factorization techniques for recommender systems}.
\newblock \bibinfo{journal}{\emph{Computer}} \bibinfo{volume}{42}, \bibinfo{number}{8} (\bibinfo{year}{2009}), \bibinfo{pages}{30--37}.
\newblock


\bibitem[Kumar et~al\mbox{.}(2024)]%
        {kumar2024training}
\bibfield{author}{\bibinfo{person}{Aviral Kumar}, \bibinfo{person}{Vincent Zhuang}, \bibinfo{person}{Rishabh Agarwal}, \bibinfo{person}{Yi Su}, \bibinfo{person}{John~D Co-Reyes}, \bibinfo{person}{Avi Singh}, \bibinfo{person}{Kate Baumli}, \bibinfo{person}{Shariq Iqbal}, \bibinfo{person}{Colton Bishop}, \bibinfo{person}{Rebecca Roelofs}, {et~al\mbox{.}}} \bibinfo{year}{2024}\natexlab{}.
\newblock \showarticletitle{Training language models to self-correct via reinforcement learning}.
\newblock \bibinfo{journal}{\emph{arXiv preprint arXiv:2409.12917}} (\bibinfo{year}{2024}).
\newblock


\bibitem[Li et~al\mbox{.}(2023)]%
        {li2023generative}
\bibfield{author}{\bibinfo{person}{Junlong Li}, \bibinfo{person}{Shichao Sun}, \bibinfo{person}{Weizhe Yuan}, \bibinfo{person}{Run-Ze Fan}, \bibinfo{person}{Hai Zhao}, {and} \bibinfo{person}{Pengfei Liu}.} \bibinfo{year}{2023}\natexlab{}.
\newblock \showarticletitle{Generative judge for evaluating alignment}.
\newblock \bibinfo{journal}{\emph{arXiv preprint arXiv:2310.05470}} (\bibinfo{year}{2023}).
\newblock


\bibitem[Li et~al\mbox{.}(2019)]%
        {li2019fi}
\bibfield{author}{\bibinfo{person}{Zekun Li}, \bibinfo{person}{Zeyu Cui}, \bibinfo{person}{Shu Wu}, \bibinfo{person}{Xiaoyu Zhang}, {and} \bibinfo{person}{Liang Wang}.} \bibinfo{year}{2019}\natexlab{}.
\newblock \showarticletitle{Fi-gnn: Modeling feature interactions via graph neural networks for ctr prediction}. In \bibinfo{booktitle}{\emph{Proceedings of the 28th ACM international conference on information and knowledge management}}. \bibinfo{pages}{539--548}.
\newblock


\bibitem[Liao et~al\mbox{.}(2023)]%
        {liao2023llara}
\bibfield{author}{\bibinfo{person}{Jiayi Liao}, \bibinfo{person}{Sihang Li}, \bibinfo{person}{Zhengyi Yang}, \bibinfo{person}{Jiancan Wu}, \bibinfo{person}{Yancheng Yuan}, \bibinfo{person}{Xiang Wang}, {and} \bibinfo{person}{Xiangnan He}.} \bibinfo{year}{2023}\natexlab{}.
\newblock \showarticletitle{Llara: Aligning large language models with sequential recommenders}.
\newblock \bibinfo{journal}{\emph{arXiv preprint arXiv:2312.02445}} (\bibinfo{year}{2023}).
\newblock


\bibitem[Lin et~al\mbox{.}(2024)]%
        {lin2024bridging}
\bibfield{author}{\bibinfo{person}{Xinyu Lin}, \bibinfo{person}{Wenjie Wang}, \bibinfo{person}{Yongqi Li}, \bibinfo{person}{Fuli Feng}, \bibinfo{person}{See-Kiong Ng}, {and} \bibinfo{person}{Tat-Seng Chua}.} \bibinfo{year}{2024}\natexlab{}.
\newblock \showarticletitle{Bridging items and language: A transition paradigm for large language model-based recommendation}. In \bibinfo{booktitle}{\emph{Proceedings of the 30th ACM SIGKDD Conference on Knowledge Discovery and Data Mining}}. \bibinfo{pages}{1816--1826}.
\newblock


\bibitem[Liu et~al\mbox{.}(2023)]%
        {liu2023chatgpt}
\bibfield{author}{\bibinfo{person}{Junling Liu}, \bibinfo{person}{Chao Liu}, \bibinfo{person}{Peilin Zhou}, \bibinfo{person}{Renjie Lv}, \bibinfo{person}{Kang Zhou}, {and} \bibinfo{person}{Yan Zhang}.} \bibinfo{year}{2023}\natexlab{}.
\newblock \showarticletitle{Is chatgpt a good recommender? a preliminary study}.
\newblock \bibinfo{journal}{\emph{arXiv preprint arXiv:2304.10149}} (\bibinfo{year}{2023}).
\newblock


\bibitem[Madaan et~al\mbox{.}(2024)]%
        {madaan2024self}
\bibfield{author}{\bibinfo{person}{Aman Madaan}, \bibinfo{person}{Niket Tandon}, \bibinfo{person}{Prakhar Gupta}, \bibinfo{person}{Skyler Hallinan}, \bibinfo{person}{Luyu Gao}, \bibinfo{person}{Sarah Wiegreffe}, \bibinfo{person}{Uri Alon}, \bibinfo{person}{Nouha Dziri}, \bibinfo{person}{Shrimai Prabhumoye}, \bibinfo{person}{Yiming Yang}, {et~al\mbox{.}}} \bibinfo{year}{2024}\natexlab{}.
\newblock \showarticletitle{Self-refine: Iterative refinement with self-feedback}.
\newblock \bibinfo{journal}{\emph{Advances in Neural Information Processing Systems}}  \bibinfo{volume}{36} (\bibinfo{year}{2024}).
\newblock


\bibitem[Olausson et~al\mbox{.}(2023)]%
        {olausson2023self}
\bibfield{author}{\bibinfo{person}{Theo~X Olausson}, \bibinfo{person}{Jeevana~Priya Inala}, \bibinfo{person}{Chenglong Wang}, \bibinfo{person}{Jianfeng Gao}, {and} \bibinfo{person}{Armando Solar-Lezama}.} \bibinfo{year}{2023}\natexlab{}.
\newblock \showarticletitle{Is Self-Repair a Silver Bullet for Code Generation?}. In \bibinfo{booktitle}{\emph{The Twelfth International Conference on Learning Representations}}.
\newblock


\bibitem[Ouyang et~al\mbox{.}(2022)]%
        {ouyang2022training}
\bibfield{author}{\bibinfo{person}{Long Ouyang}, \bibinfo{person}{Jeffrey Wu}, \bibinfo{person}{Xu Jiang}, \bibinfo{person}{Diogo Almeida}, \bibinfo{person}{Carroll Wainwright}, \bibinfo{person}{Pamela Mishkin}, \bibinfo{person}{Chong Zhang}, \bibinfo{person}{Sandhini Agarwal}, \bibinfo{person}{Katarina Slama}, \bibinfo{person}{Alex Ray}, {et~al\mbox{.}}} \bibinfo{year}{2022}\natexlab{}.
\newblock \showarticletitle{Training language models to follow instructions with human feedback}.
\newblock \bibinfo{journal}{\emph{Advances in neural information processing systems}}  \bibinfo{volume}{35} (\bibinfo{year}{2022}), \bibinfo{pages}{27730--27744}.
\newblock


\bibitem[Pan et~al\mbox{.}(2023)]%
        {pan2023logic}
\bibfield{author}{\bibinfo{person}{Liangming Pan}, \bibinfo{person}{Alon Albalak}, \bibinfo{person}{Xinyi Wang}, {and} \bibinfo{person}{William~Yang Wang}.} \bibinfo{year}{2023}\natexlab{}.
\newblock \showarticletitle{Logic-lm: Empowering large language models with symbolic solvers for faithful logical reasoning}.
\newblock \bibinfo{journal}{\emph{arXiv preprint arXiv:2305.12295}} (\bibinfo{year}{2023}).
\newblock


\bibitem[Paul et~al\mbox{.}(2023)]%
        {paul2023refiner}
\bibfield{author}{\bibinfo{person}{Debjit Paul}, \bibinfo{person}{Mete Ismayilzada}, \bibinfo{person}{Maxime Peyrard}, \bibinfo{person}{Beatriz Borges}, \bibinfo{person}{Antoine Bosselut}, \bibinfo{person}{Robert West}, {and} \bibinfo{person}{Boi Faltings}.} \bibinfo{year}{2023}\natexlab{}.
\newblock \showarticletitle{Refiner: Reasoning feedback on intermediate representations}.
\newblock \bibinfo{journal}{\emph{arXiv preprint arXiv:2304.01904}} (\bibinfo{year}{2023}).
\newblock


\bibitem[Qu et~al\mbox{.}(2024)]%
        {qu2024recursive}
\bibfield{author}{\bibinfo{person}{Yuxiao Qu}, \bibinfo{person}{Tianjun Zhang}, \bibinfo{person}{Naman Garg}, {and} \bibinfo{person}{Aviral Kumar}.} \bibinfo{year}{2024}\natexlab{}.
\newblock \showarticletitle{Recursive introspection: Teaching language model agents how to self-improve}.
\newblock \bibinfo{journal}{\emph{arXiv preprint arXiv:2407.18219}} (\bibinfo{year}{2024}).
\newblock


\bibitem[Shinn et~al\mbox{.}(2024)]%
        {shinn2024reflexion}
\bibfield{author}{\bibinfo{person}{Noah Shinn}, \bibinfo{person}{Federico Cassano}, \bibinfo{person}{Ashwin Gopinath}, \bibinfo{person}{Karthik Narasimhan}, {and} \bibinfo{person}{Shunyu Yao}.} \bibinfo{year}{2024}\natexlab{}.
\newblock \showarticletitle{Reflexion: Language agents with verbal reinforcement learning}.
\newblock \bibinfo{journal}{\emph{Advances in Neural Information Processing Systems}}  \bibinfo{volume}{36} (\bibinfo{year}{2024}).
\newblock


\bibitem[Snell et~al\mbox{.}(2024)]%
        {snell2024scaling}
\bibfield{author}{\bibinfo{person}{Charlie Snell}, \bibinfo{person}{Jaehoon Lee}, \bibinfo{person}{Kelvin Xu}, {and} \bibinfo{person}{Aviral Kumar}.} \bibinfo{year}{2024}\natexlab{}.
\newblock \showarticletitle{Scaling llm test-time compute optimally can be more effective than scaling model parameters}.
\newblock \bibinfo{journal}{\emph{arXiv preprint arXiv:2408.03314}} (\bibinfo{year}{2024}).
\newblock


\bibitem[Song et~al\mbox{.}(2019)]%
        {song2019autoint}
\bibfield{author}{\bibinfo{person}{Weiping Song}, \bibinfo{person}{Chence Shi}, \bibinfo{person}{Zhiping Xiao}, \bibinfo{person}{Zhijian Duan}, \bibinfo{person}{Yewen Xu}, \bibinfo{person}{Ming Zhang}, {and} \bibinfo{person}{Jian Tang}.} \bibinfo{year}{2019}\natexlab{}.
\newblock \showarticletitle{Autoint: Automatic feature interaction learning via self-attentive neural networks}. In \bibinfo{booktitle}{\emph{Proceedings of the 28th ACM international conference on information and knowledge management}}. \bibinfo{pages}{1161--1170}.
\newblock


\bibitem[Taori et~al\mbox{.}(2023)]%
        {taori2023stanford}
\bibfield{author}{\bibinfo{person}{Rohan Taori}, \bibinfo{person}{Ishaan Gulrajani}, \bibinfo{person}{Tianyi Zhang}, \bibinfo{person}{Yann Dubois}, \bibinfo{person}{Xuechen Li}, \bibinfo{person}{Carlos Guestrin}, \bibinfo{person}{Percy Liang}, {and} \bibinfo{person}{Tatsunori~B Hashimoto}.} \bibinfo{year}{2023}\natexlab{}.
\newblock \showarticletitle{Stanford alpaca: an instruction-following llama model (2023)}.
\newblock \bibinfo{journal}{\emph{URL https://github. com/tatsu-lab/stanford\_alpaca}} \bibinfo{volume}{1}, \bibinfo{number}{9} (\bibinfo{year}{2023}).
\newblock


\bibitem[Touvron et~al\mbox{.}(2023)]%
        {touvron2023llama}
\bibfield{author}{\bibinfo{person}{Hugo Touvron}, \bibinfo{person}{Thibaut Lavril}, \bibinfo{person}{Gautier Izacard}, \bibinfo{person}{Xavier Martinet}, \bibinfo{person}{Marie-Anne Lachaux}, \bibinfo{person}{Timoth{\'e}e Lacroix}, \bibinfo{person}{Baptiste Rozi{\`e}re}, \bibinfo{person}{Naman Goyal}, \bibinfo{person}{Eric Hambro}, \bibinfo{person}{Faisal Azhar}, {et~al\mbox{.}}} \bibinfo{year}{2023}\natexlab{}.
\newblock \showarticletitle{Llama: Open and efficient foundation language models}.
\newblock \bibinfo{journal}{\emph{arXiv preprint arXiv:2302.13971}} (\bibinfo{year}{2023}).
\newblock


\bibitem[Wang et~al\mbox{.}(2024b)]%
        {wang2024rethinking}
\bibfield{author}{\bibinfo{person}{Hanbing Wang}, \bibinfo{person}{Xiaorui Liu}, \bibinfo{person}{Wenqi Fan}, \bibinfo{person}{Xiangyu Zhao}, \bibinfo{person}{Venkataramana Kini}, \bibinfo{person}{Devendra Yadav}, \bibinfo{person}{Fei Wang}, \bibinfo{person}{Zhen Wen}, \bibinfo{person}{Jiliang Tang}, {and} \bibinfo{person}{Hui Liu}.} \bibinfo{year}{2024}\natexlab{b}.
\newblock \showarticletitle{Rethinking large language model architectures for sequential recommendations}.
\newblock \bibinfo{journal}{\emph{arXiv preprint arXiv:2402.09543}} (\bibinfo{year}{2024}).
\newblock


\bibitem[Wang and Lim(2023)]%
        {wang2023zero}
\bibfield{author}{\bibinfo{person}{Lei Wang} {and} \bibinfo{person}{Ee-Peng Lim}.} \bibinfo{year}{2023}\natexlab{}.
\newblock \showarticletitle{Zero-shot next-item recommendation using large pretrained language models}.
\newblock \bibinfo{journal}{\emph{arXiv preprint arXiv:2304.03153}} (\bibinfo{year}{2023}).
\newblock


\bibitem[Wang et~al\mbox{.}(2017)]%
        {wang2017deep}
\bibfield{author}{\bibinfo{person}{Ruoxi Wang}, \bibinfo{person}{Bin Fu}, \bibinfo{person}{Gang Fu}, {and} \bibinfo{person}{Mingliang Wang}.} \bibinfo{year}{2017}\natexlab{}.
\newblock \showarticletitle{Deep \& cross network for ad click predictions}.
\newblock In \bibinfo{booktitle}{\emph{Proceedings of the ADKDD'17}}. \bibinfo{pages}{1--7}.
\newblock


\bibitem[Wang et~al\mbox{.}(2024a)]%
        {wang2024learnable}
\bibfield{author}{\bibinfo{person}{Wenjie Wang}, \bibinfo{person}{Honghui Bao}, \bibinfo{person}{Xinyu Lin}, \bibinfo{person}{Jizhi Zhang}, \bibinfo{person}{Yongqi Li}, \bibinfo{person}{Fuli Feng}, \bibinfo{person}{See-Kiong Ng}, {and} \bibinfo{person}{Tat-Seng Chua}.} \bibinfo{year}{2024}\natexlab{a}.
\newblock \showarticletitle{Learnable Tokenizer for LLM-based Generative Recommendation}.
\newblock \bibinfo{journal}{\emph{arXiv preprint arXiv:2405.07314}} (\bibinfo{year}{2024}).
\newblock


\bibitem[Wang et~al\mbox{.}(2022)]%
        {wang2022self}
\bibfield{author}{\bibinfo{person}{Xuezhi Wang}, \bibinfo{person}{Jason Wei}, \bibinfo{person}{Dale Schuurmans}, \bibinfo{person}{Quoc Le}, \bibinfo{person}{Ed Chi}, \bibinfo{person}{Sharan Narang}, \bibinfo{person}{Aakanksha Chowdhery}, {and} \bibinfo{person}{Denny Zhou}.} \bibinfo{year}{2022}\natexlab{}.
\newblock \showarticletitle{Self-consistency improves chain of thought reasoning in language models}.
\newblock \bibinfo{journal}{\emph{arXiv preprint arXiv:2203.11171}} (\bibinfo{year}{2022}).
\newblock


\bibitem[Welleck et~al\mbox{.}(2022)]%
        {welleck2022generating}
\bibfield{author}{\bibinfo{person}{Sean Welleck}, \bibinfo{person}{Ximing Lu}, \bibinfo{person}{Peter West}, \bibinfo{person}{Faeze Brahman}, \bibinfo{person}{Tianxiao Shen}, \bibinfo{person}{Daniel Khashabi}, {and} \bibinfo{person}{Yejin Choi}.} \bibinfo{year}{2022}\natexlab{}.
\newblock \showarticletitle{Generating sequences by learning to self-correct}.
\newblock \bibinfo{journal}{\emph{arXiv preprint arXiv:2211.00053}} (\bibinfo{year}{2022}).
\newblock


\bibitem[Wu et~al\mbox{.}(2024)]%
        {wu2024large}
\bibfield{author}{\bibinfo{person}{Zhenyu Wu}, \bibinfo{person}{Qingkai Zeng}, \bibinfo{person}{Zhihan Zhang}, \bibinfo{person}{Zhaoxuan Tan}, \bibinfo{person}{Chao Shen}, {and} \bibinfo{person}{Meng Jiang}.} \bibinfo{year}{2024}\natexlab{}.
\newblock \showarticletitle{Large language models can self-correct with key condition verification}. In \bibinfo{booktitle}{\emph{Proceedings of the 2024 Conference on Empirical Methods in Natural Language Processing}}. \bibinfo{pages}{12846--12867}.
\newblock


\bibitem[Xi et~al\mbox{.}(2024a)]%
        {xi2024towards}
\bibfield{author}{\bibinfo{person}{Yunjia Xi}, \bibinfo{person}{Weiwen Liu}, \bibinfo{person}{Jianghao Lin}, \bibinfo{person}{Xiaoling Cai}, \bibinfo{person}{Hong Zhu}, \bibinfo{person}{Jieming Zhu}, \bibinfo{person}{Bo Chen}, \bibinfo{person}{Ruiming Tang}, \bibinfo{person}{Weinan Zhang}, {and} \bibinfo{person}{Yong Yu}.} \bibinfo{year}{2024}\natexlab{a}.
\newblock \showarticletitle{Towards open-world recommendation with knowledge augmentation from large language models}. In \bibinfo{booktitle}{\emph{Proceedings of the 18th ACM Conference on Recommender Systems}}. \bibinfo{pages}{12--22}.
\newblock


\bibitem[Xi et~al\mbox{.}(2024b)]%
        {xi2024enhancing}
\bibfield{author}{\bibinfo{person}{Zhiheng Xi}, \bibinfo{person}{Dingwen Yang}, \bibinfo{person}{Jixuan Huang}, \bibinfo{person}{Jiafu Tang}, \bibinfo{person}{Guanyu Li}, \bibinfo{person}{Yiwen Ding}, \bibinfo{person}{Wei He}, \bibinfo{person}{Boyang Hong}, \bibinfo{person}{Shihan Do}, \bibinfo{person}{Wenyu Zhan}, {et~al\mbox{.}}} \bibinfo{year}{2024}\natexlab{b}.
\newblock \showarticletitle{Enhancing LLM Reasoning via Critique Models with Test-Time and Training-Time Supervision}.
\newblock \bibinfo{journal}{\emph{arXiv preprint arXiv:2411.16579}} (\bibinfo{year}{2024}).
\newblock


\bibitem[Xia et~al\mbox{.}(2025)]%
        {xia2025hierarchical}
\bibfield{author}{\bibinfo{person}{Yu Xia}, \bibinfo{person}{Rui Zhong}, \bibinfo{person}{Hao Gu}, \bibinfo{person}{Wei Yang}, \bibinfo{person}{Chi Lu}, \bibinfo{person}{Peng Jiang}, {and} \bibinfo{person}{Kun Gai}.} \bibinfo{year}{2025}\natexlab{}.
\newblock \showarticletitle{Hierarchical Tree Search-based User Lifelong Behavior Modeling on Large Language Model}.
\newblock \bibinfo{journal}{\emph{arXiv preprint arXiv:2505.19505}} (\bibinfo{year}{2025}).
\newblock


\bibitem[Xu et~al\mbox{.}(2024)]%
        {xu2024prompting}
\bibfield{author}{\bibinfo{person}{Lanling Xu}, \bibinfo{person}{Junjie Zhang}, \bibinfo{person}{Bingqian Li}, \bibinfo{person}{Jinpeng Wang}, \bibinfo{person}{Mingchen Cai}, \bibinfo{person}{Wayne~Xin Zhao}, {and} \bibinfo{person}{Ji-Rong Wen}.} \bibinfo{year}{2024}\natexlab{}.
\newblock \showarticletitle{Prompting large language models for recommender systems: A comprehensive framework and empirical analysis}.
\newblock \bibinfo{journal}{\emph{arXiv preprint arXiv:2401.04997}} (\bibinfo{year}{2024}).
\newblock


\bibitem[Yang et~al\mbox{.}(2024)]%
        {yang2024qwen2}
\bibfield{author}{\bibinfo{person}{An Yang}, \bibinfo{person}{Baosong Yang}, \bibinfo{person}{Beichen Zhang}, \bibinfo{person}{Binyuan Hui}, \bibinfo{person}{Bo Zheng}, \bibinfo{person}{Bowen Yu}, \bibinfo{person}{Chengyuan Li}, \bibinfo{person}{Dayiheng Liu}, \bibinfo{person}{Fei Huang}, \bibinfo{person}{Haoran Wei}, {et~al\mbox{.}}} \bibinfo{year}{2024}\natexlab{}.
\newblock \showarticletitle{Qwen2. 5 Technical Report}.
\newblock \bibinfo{journal}{\emph{arXiv preprint arXiv:2412.15115}} (\bibinfo{year}{2024}).
\newblock


\bibitem[Yu et~al\mbox{.}(2023)]%
        {yu2023improving}
\bibfield{author}{\bibinfo{person}{Wenhao Yu}, \bibinfo{person}{Zhihan Zhang}, \bibinfo{person}{Zhenwen Liang}, \bibinfo{person}{Meng Jiang}, {and} \bibinfo{person}{Ashish Sabharwal}.} \bibinfo{year}{2023}\natexlab{}.
\newblock \showarticletitle{Improving language models via plug-and-play retrieval feedback}.
\newblock \bibinfo{journal}{\emph{arXiv preprint arXiv:2305.14002}} (\bibinfo{year}{2023}).
\newblock


\bibitem[Zhang et~al\mbox{.}(2023)]%
        {zhang2023recommendation}
\bibfield{author}{\bibinfo{person}{Junjie Zhang}, \bibinfo{person}{Ruobing Xie}, \bibinfo{person}{Yupeng Hou}, \bibinfo{person}{Xin Zhao}, \bibinfo{person}{Leyu Lin}, {and} \bibinfo{person}{Ji-Rong Wen}.} \bibinfo{year}{2023}\natexlab{}.
\newblock \showarticletitle{Recommendation as instruction following: A large language model empowered recommendation approach}.
\newblock \bibinfo{journal}{\emph{ACM Transactions on Information Systems}} (\bibinfo{year}{2023}).
\newblock


\bibitem[Zhao et~al\mbox{.}(2023)]%
        {zhao2023recommender}
\bibfield{author}{\bibinfo{person}{Zihuai Zhao}, \bibinfo{person}{Wenqi Fan}, \bibinfo{person}{Jiatong Li}, \bibinfo{person}{Yunqing Liu}, \bibinfo{person}{Xiaowei Mei}, \bibinfo{person}{Yiqi Wang}, \bibinfo{person}{Zhen Wen}, \bibinfo{person}{Fei Wang}, \bibinfo{person}{Xiangyu Zhao}, \bibinfo{person}{Jiliang Tang}, {et~al\mbox{.}}} \bibinfo{year}{2023}\natexlab{}.
\newblock \showarticletitle{Recommender systems in the era of large language models (llms)}.
\newblock \bibinfo{journal}{\emph{arXiv preprint arXiv:2307.02046}} (\bibinfo{year}{2023}).
\newblock


\bibitem[Zheng et~al\mbox{.}(2024)]%
        {zheng2024critic}
\bibfield{author}{\bibinfo{person}{Xin Zheng}, \bibinfo{person}{Jie Lou}, \bibinfo{person}{Boxi Cao}, \bibinfo{person}{Xueru Wen}, \bibinfo{person}{Yuqiu Ji}, \bibinfo{person}{Hongyu Lin}, \bibinfo{person}{Yaojie Lu}, \bibinfo{person}{Xianpei Han}, \bibinfo{person}{Debing Zhang}, {and} \bibinfo{person}{Le Sun}.} \bibinfo{year}{2024}\natexlab{}.
\newblock \showarticletitle{Critic-cot: Boosting the reasoning abilities of large language model via chain-of-thoughts critic}.
\newblock \bibinfo{journal}{\emph{arXiv preprint arXiv:2408.16326}} (\bibinfo{year}{2024}).
\newblock


\bibitem[Zhou et~al\mbox{.}(2019)]%
        {zhou2019deep}
\bibfield{author}{\bibinfo{person}{Guorui Zhou}, \bibinfo{person}{Na Mou}, \bibinfo{person}{Ying Fan}, \bibinfo{person}{Qi Pi}, \bibinfo{person}{Weijie Bian}, \bibinfo{person}{Chang Zhou}, \bibinfo{person}{Xiaoqiang Zhu}, {and} \bibinfo{person}{Kun Gai}.} \bibinfo{year}{2019}\natexlab{}.
\newblock \showarticletitle{Deep interest evolution network for click-through rate prediction}. In \bibinfo{booktitle}{\emph{Proceedings of the AAAI conference on artificial intelligence}}, Vol.~\bibinfo{volume}{33}. \bibinfo{pages}{5941--5948}.
\newblock


\end{thebibliography}

\appendix

\end{document}